\documentclass[aps, prd, showkeys,superscriptaddress, nofootinbib,floatfix]{revtex4-2}

\usepackage{graphicx}
\usepackage{dcolumn}
\usepackage{bm}
\usepackage{amsmath,amstext}
\usepackage[T1]{fontenc}

\usepackage{mathtools}
\usepackage[bookmarksnumbered, pdfpagelabels=true, plainpages=false, colorlinks=true, linkcolor=blue, citecolor=blue, urlcolor=blue]{hyperref}

\usepackage{xcolor}

\begin{document}

\title{Acausality-driven instabilities in transient relativistic viscous hydrodynamics}

\author{Lorenzo Gavassino}
\email{lorenzo.gavassino@vanderbilt.edu}
\affiliation{Department of Mathematics, Vanderbilt University, Nashville TN 37240}

\author{Henry Hirvonen}
\email{henry.hirvonen@vanderbilt.edu}
\affiliation{Department of Mathematics, Vanderbilt University, Nashville TN 37240}

\author{Jean-François Paquet}
\email{jean-francois.paquet@vanderbilt.edu}
\affiliation{Department of Physics and Astronomy, Vanderbilt University, Nashville TN 37240}
\affiliation{Department of Mathematics, Vanderbilt University, Nashville TN 37240}

\author{Mayank Singh}
\email{mayank.singh@vanderbilt.edu}
\affiliation{Department of Physics and Astronomy, Vanderbilt University, Nashville TN 37240}

\author{Gabriel Soares Rocha}
\email{gabrielsr@id.uff.br}
\affiliation{Instituto de F\'{\i}sica, Universidade Federal Fluminense, Niter\'{o}i, Rio de Janeiro, 24210-346, Brazil}
\affiliation{Department of Physics and Astronomy, Vanderbilt University, Nashville TN 37240}

\begin{abstract}
We investigate non-linear instabilities stemming from superluminal propagation of information in Israel-Stewart-like models of relativistic viscous fluid dynamics. In relativity, the characteristic speed of propagation of information, $w$, and the speed of the fluid, $v$, allow us to differentiate between regimes of the hydrodynamic equations that are acausal but stable ($w>1$), unstable ($v^{2} w^{2} \geq 1$), and covariantly ill-posed ($w^{2} \leq 0$). As an analytical benchmark, we present a new solution that illustrates these distinct regimes.
We compare this analytical solution to the result of a numerical relativistic viscous fluid dynamics solver, and confirm that the analytical result can be recovered numerically in the stable regime, whether causal or acausal. The onset of numerical instabilities is further found to occur in the regime predicted by the analytical solution.    
\end{abstract}

\maketitle

\section{Introduction}

The relativistic generalization of the Navier-Stokes equations originally developed by \citet{Eckart40} and \citet{landau6} is known to be plagued by linear instabilities \cite{Hishcock1983} unless the equations are solved under strong symmetry constraints.
An alternative formulation\footnote{Other formulations include the Bemfica-Disconzi-Noronha-Kovtun (BDNK) theory \cite{Bemfica:2017wps, Bemfica:2019knx, Bemfica:2020zjp, Kovtun:2019hdm}, and Density-frame hydrodynamics \cite{Bhambure:2024gnf,Bhambure:2024axa}. See also refs.~\cite{Jeon:2015dfa,Rocha:2023ilf}, for discussions on hydrodynamic formulations in relativity.} of relativistic viscous hydrodynamics aimed at addressing these instabilities is the Israel-Stewart \cite{Israel:1976tn,Israel:1979wp} (IS) theory, which is the relativistic generalization of Maxwell's model of viscoelasticity \cite[\S 36]{landau7}. The central idea behind the Israel-Stewart theory, and behind all ``transient theories'', is to modify the Navier-Stokes constitutive relations by adding a delay in the response of the viscous stress component (e.g. the bulk pressure $\Pi$) to the corresponding strain rate (e.g. the expansion rate $\theta=\partial_\mu u^\mu$, with $u^\mu$ the flow velocity), with a characteristic delay timescale (e.g. the bulk relaxation time $\tau_\Pi$). Explicitly,
\begin{flalign}
\text{Navier-Stokes:} && \label{NS} \Pi & \propto - \theta \;,&\\
\text{Transient hydrodynamics:} && \label{IS-int} \tau_\Pi u^\mu \partial_\mu \Pi +\Pi &\propto - \theta \;.
\end{flalign}

Equation~\eqref{IS-int} was later refined by many authors \cite{Baier:2007ix,Denicol:2012cn,Bhattacharyya:2007vjd}, who, following power counting procedures \cite{Rocha:2023ilf,Denicol:2021,Romatschke:2017ejr}, added other second-order terms (e.g. terms of the form $\Pi^2$, $\Pi \theta$, and $\theta^2$) on the right-hand side of eq. \eqref{IS-int}, leading to the predominant formulation that is currently used in simulations of ultrarelativistic heavy-ion collisions~\cite{Romatschke:2017ejr,Jeon:2015dfa,DerradideSouza:2015kpt,Gale:2013da}. 

One feature that made Israel-Stewart-type frameworks attractive to the relativity community was their causal behavior. In fact, for many years, it was believed that information always traveled at subluminal speeds within Israel-Stewart theory, provided that the transport coefficients are chosen appropriately. This belief arose from a sequence of works \cite{Hishcock1983,OlsonLifsh1990,Pu:2009fj}, where the characteristic speeds of propagation of Israel-Stewart were shown to be smaller than the speed of light \emph{near equilibrium} (i.e. for small viscous stresses), if the ratio $\zeta/\tau_\Pi$ was taken to be small enough. However, the belief that Israel-Stewart theory is \textit{always} causal was recently proven to be false in refs.~\cite{Bemfica:2019cop,Bemfica:2020xym}. There, the causality analysis was carried out on states arbitrarily far from equilibrium, and it was found that the speed of information is sensitive to the intensity of the viscous stresses. In particular, when the latter are very large, causality can always be violated, independently from the particular choice of $\zeta$ and $\tau_\Pi$. For example, in a fluid at zero chemical potential, with only bulk viscosity, the speed of propagation of information $w$ (in the local rest frame) is given by
\begin{equation}\label{bulkione}
w^2=c_s^2 +\dfrac{\zeta}{\tau_\Pi (\varepsilon{+}P{+}\Pi)} \, ,
\end{equation}
where $\varepsilon$, $P$, and $c_s^2 = \partial P/\partial \varepsilon$ are respectively the energy density, pressure, and speed of sound squared of the fluid.
As it can be seen above, if $\Pi \rightarrow -(\varepsilon{+}P)$, causality is necessarily violated\footnote{This statement would no longer be true for e.g.~BDNK theory~\cite{Bemfica:2017wps, Bemfica:2019knx, Bemfica:2020zjp, Kovtun:2019hdm}, where the characteristic propagation speeds depend only on the transport coefficients (e.g.~$\zeta$) and not on the stresses (e.g.~$\Pi$).}, for any choice of $\zeta/\tau_\Pi > 0$.

When the Israel-Stewart theory is solved numerically to study relativistic viscous fluids, such as the quark-gluon plasma produced in high-energy nuclear collisions, one cannot control whether the conditions for superluminal propagation of information occur~\cite{Plumberg:2021bme,Chiu:2021muk,ExTrEMe:2023nhy,Domingues:2024pom}. While initial conditions that directly lead to acausal propagation can be excluded to limit acausal behavior~\cite{Domingues:2024pom}, acausalities can emerge at any point later in the evolution~\cite{Hoshino:2024qun}. Importantly, the consequences of acausality are subtle; it is not atypical for causality to be broken when degrees of freedom are removed from an underlying causal formalism such as kinetic theory (see e.g. \cite{GavassinoDispersions2024}). 
On the other hand, the possible existence of acausality-driven instabilities poses potentially serious challenges for simulations. They can lead to the rapid accumulation of errors on the numerical solution with respect to the `true' solution of the partial differential equations. Alternatively, instabilities can be so severe that the partial differential equations are ill-posed. The goal of this work is to investigate the interplay between acausality-driven instabilities and the numerical solution of relativistic dissipative hydrodynamic equations of motion. In Section \ref{sec:th-sumry}, we summarize the connection between acausality and instability, and derive the condition $v^{2} w^{2} \geq 1$, where $v$ is the magnitude of the three-velocity, for a system to be acausality-driven unstable. Section \ref{sec:simple_test} forms the core of the present work. There, we define the causality-stability classification of a given system as (i) causal, (ii) acausal-stable and (iii) unstable, which we will refer to at various points as `good', `bad' or `ugly', respectively, for simplicity.  As our main result, we introduce a new benchmark solution for which the evolution of the hydrodynamic fields is known analytically, and compare it with the numerical solution provided by the solver MUSIC \cite{Schenke:2010nt,Schenke:2010rr,Paquet:2015lta} using initial conditions in the various causality-stability regimes. In Section \ref{eq:GBU-test-2D}, as a first step to generalizing this work, we present $(2+1)$-dimensional numerical tests with a type of initial conditions used in heavy-ion collisions, but with simpler equations of motion than typically used in the field.
Section~\ref{sec:concl} summarizes our findings. Appendix \ref{sec:hom-fluid-obed} assesses the obedience to local conservation laws of the numerical solution in the different causality-stability regimes and in Appendix \ref{apn:details-IC}, we provide more details on initial state prescriptions in numerical relativistic hydrodynamic solvers and outline an alternative prescription to be employed in such numerical solutions.

%%%%%%%%%%%%%%%%%%%%%%%%%%%%%%%%%%%%%%%%%%%%%%%%%%%%%%%%%%%%
%%%%%%%%%%%%%%%%%%%%%%%%%%%%%%%%%%%%%%%%%%%%%%%%%%%%%%%%%%%%
\section{Link between acausality and instability: Theoretical summary}
\label{sec:th-sumry}
%%%%%%%%%%%%%%%%%%%%%%%%%%%%%%%%%%%%%%%%%%%%%%%%%%%%%%%%%%%%
%%%%%%%%%%%%%%%%%%%%%%%%%%%%%%%%%%%%%%%%%%%%%%%%%%%%%%%%%%%%

The existence of a connection between causality violations and instabilities in relativistic hydrodynamics has been observed several times in the past \cite{Hishcock1983,Israel_2009_inbook,Pu:2009fj,Bemfica:2020zjp}. However, its physical origin was clarified only recently \cite{Gavassino:2021kjm,Gavassino:2021owo}. In this section, we briefly outline the modern view on the subject.

%
%%%%%%%%%%%%%%%%%%%%%%%%%%%%%%%%%%%%%%%%%%%%%%%%%%%%%%%%%%%%
%%%%%%%%%%%%%%%%%%%%%%%%%%%%%%%%%%%%%%%%%%%%%%%%%%%%%%%%%%%%
\vspace{-0.2cm}
\subsection{Backward dissipation (hyperbolic case)}\label{hy1}
\vspace{-0.2cm}
%%%%%%%%%%%%%%%%%%%%%%%%%%%%%%%%%%%%%%%%%%%%%%%%%%%%%%%%%%%%
%%%%%%%%%%%%%%%%%%%%%%%%%%%%%%%%%%%%%%%%%%%%%%%%%%%%%%%%%%%%

Under normal conditions, the equations of motion of Israel-Stewart theory are hyperbolic: if we perturb the fluid at a given point $x^\mu$ in spacetime, the information about our intervention travels inside a (possibly distorted) conical shape, which is known as the ``outermost characteristic cone'' \cite{CourantHilbert2_book}, or ``acoustic cone'' (in analogy with the lightcone).\footnote{The acoustic cone is a characteristic manifold, along which non-linear wavefronts propagate. As such, it can be computed using characteristic manifold techniques for partial differential equations \cite{Bemfica:2019cop,Bemfica:2020xym}. The calculation requires only the knowledge of the coefficients of the terms with the highest derivative orders. The case of pure bulk (but with non-zero chemical potential) is discussed in ref.~\cite{Bemfica:2019cop}.} In the case of a fluid with only bulk viscous corrections, the tip of the acoustic cone at $x^\mu$ is isotropic in the local rest frame of the fluid, and its opening is set by the speed of propagation of information, $w\geq 0$, in eq. \eqref{bulkione}.
We can thus identify the acoustic cone of bulk-viscous fluids with the null cone of an ``acoustic metric'' \cite{ChristodoluouShocks,Bemfica:2019cop,Babichev:2007dw,DisconziAcoustic2019},
\begin{equation}
G_{\mu \nu}=-u_\mu u_\nu +\dfrac{1}{w^2} (g_{\mu \nu}+u_\mu u_\nu) \quad \xrightarrow[]{\text{Rest frame}} \quad \dfrac{1}{w^2} \begin{bmatrix}
-w^2 & 0 & 0 & 0 \\
0 & 1 & 0 & 0 \\
0 & 0 & 1 & 0 \\
0 & 0 & 0 & 1 \\
\end{bmatrix} \, ,
\end{equation}
which is tilted along the direction of motion in a generic local Lorentz frame:
\begin{equation}
G_{\mu \nu} \quad \xrightarrow[\text{has velocity }u^\mu=(\gamma, \gamma v,0,0)]{\text{Lorentz frame where the fluid}} \quad  \dfrac{1}{w^2} 
\begin{bmatrix}
\gamma^2 (v^2{-}w^2) & -\gamma^2 (1{-}w^2) v & 0 & 0 \\
-\gamma^2 (1{-}w^2) v & \gamma^2 (1{-}v^2 w^2) & 0 & 0 \\
0 & 0 & 1 & 0 \\
0 & 0 & 0 & 1 \\
\end{bmatrix} \, .
\end{equation}
In the case where shear viscous corrections are also present, the acoustic cone is ``deformed'' already in the rest frame, since the speed of information depends on the direction of propagation of the signal relative to the principal axes of the stress tensor. Indeed, the very concept of an acoustic metric is not guaranteed to exist in the presence of shear viscosity, since the section of the acoustic cone may not be a perfect ellipsoid in general. 

\begin{figure}[t]
    \centering
\includegraphics[width=0.49\linewidth]{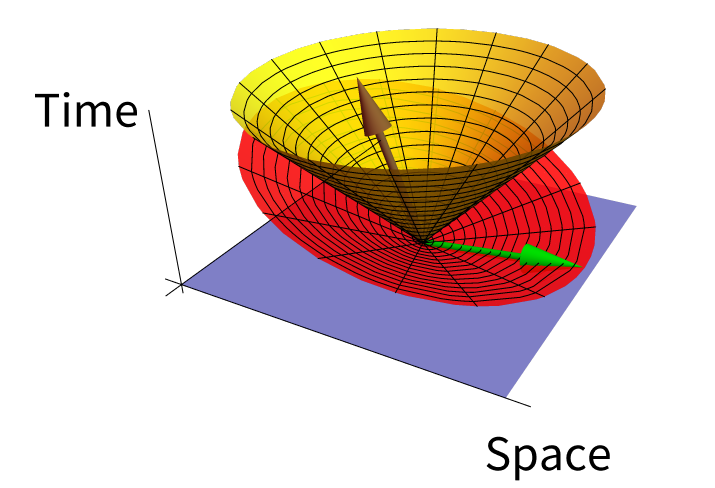}
\includegraphics[width=0.49\linewidth]{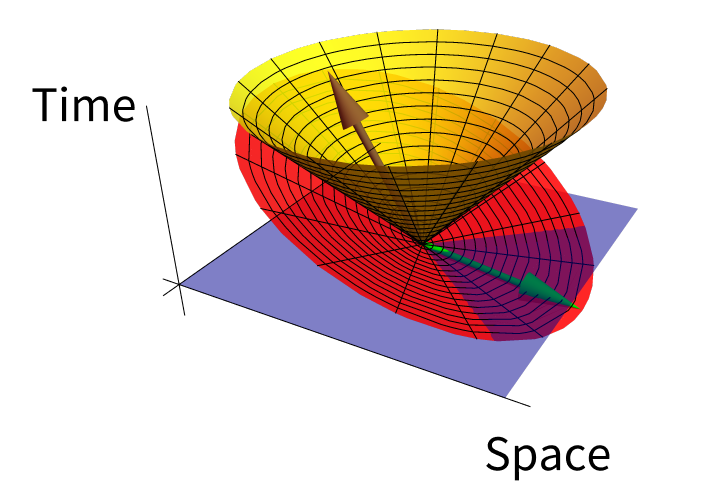}
\caption{Spacetime diagrams illustrating the difference between the acausal but stable regime (left panel) and the unstable regime (right panel). The red cone is the acoustic cone in tangent space, which is larger than the lightcone (yellow) due to acausality. In the stable case, signals (green) propagate to the future, and dissipation works as usual. In the unstable case, some signals (green) travel to the past, and they experience reversed dissipation, which causes non-equilibrium displacements to spontaneously blow up. For illustration, we took $w=2$ (working with bulk viscosity alone), and a fluid velocity $v=0.4$ in the stable case, and $0.6$ in the unstable case. The brown arrows represent the four-velocity vectors of the fluid.}
    \label{fig:vwcones}
\end{figure}

Given the above, it is clear that ``causality'' in fluid dynamics is just the statement that the acoustic cone must be contained inside the lightcone, which for bulk viscosity alone entails $w\leq 1$. Now, suppose that we violate this condition. Then, the acoustic cone becomes larger than the lightcone, and we have two possible scenarios, see fig. \ref{fig:vwcones}.
In one scenario (left panel), the acoustic cone exits the lightcone, but still points entirely to the future in the reference frame where we are solving the equations (e.g., the reference frame where the simulation is carried out). In this case, the causality violations do not give rise to instabilities, since all signals (exemplified by the green arrow) propagate from past to future, and dissipation tends to equilibrate the system. In the second scenario (right panel), part of the acoustic cone points towards the past in the reference frame where we are solving the equations. In that case, most initial conditions set on a constant-time hypersurface will lead to meaningless dynamics. In fact, signals traveling from the future to the past (green arrow) undergo \textit{reversed dissipation}, i.e., they evolve away from local equilibrium, rather than towards it \cite{Gavassino:2021owo}. As a result, any form of noise or numerical error triggers enormous departures from what would be a sensible solution if we were solving the equations in a reference frame where the cone points entirely towards the future.

Let us now determine under which conditions we enter the unstable scenario, under the assumption of a fluid with bulk viscosity alone. To this end, we note that we are in the right panel of fig. \ref{fig:vwcones} if and only if there exists a direction in space $n^\mu=(0,n^j)$, with $n^j n_j=1$, such that $G_{\mu \nu}n^\mu n^\nu=G_{jk}n^j n^k=0$. Denoting by $v$ the speed of the fluid and by $\cos\theta$ the angle between $u^j$ and $n^j$, we then obtain
\begin{equation}
\dfrac{v^2(w^2-1)}{1-v^2} =\dfrac{1}{\cos^2 \theta} \geq 1 \, .
\end{equation}
Reorganizing the above inequality, we obtain $v^2w^2\geq 1$. Thus, in order to have an instability, it is \emph{not} enough to violate causality (i.e. $w>1$): we also need the fluid to move very fast, so that also the product $v w$ goes above 1.

%%%%%%%%%%%%%%%%%%%%%%%%%%%%%%%%%%%%%%%%%%%%%%%%%%%%%%%%%%%%
%%%%%%%%%%%%%%%%%%%%%%%%%%%%%%%%%%%%%%%%%%%%%%%%%%%%%%%%%%%%
\subsection{Transversal ill-posedness (hyperbolic case)}\label{hy2}
%%%%%%%%%%%%%%%%%%%%%%%%%%%%%%%%%%%%%%%%%%%%%%%%%%%%%%%%%%%%
%%%%%%%%%%%%%%%%%%%%%%%%%%%%%%%%%%%%%%%%%%%%%%%%%%%%%%%%%%%%

It is important to note that, besides instabilities due to reversed dissipation, the fact that $vw> 1$  also leads to numerical instabilities related to the mathematical \textit{ill-posedness} of the initial value problem. The mechanism at work is analogous to what happens for ideal fluids when the speed of sound (which in the ideal case is the characteristic speed of propagation of information, $w = c_{s}$) exceeds that of light, provided also that the number of spatial dimensions is greater than 1. This can be seen from the following general argument (see also \cite{Adams:2006sv,Babichev:2007dw}).

Fix a spacetime event $x^\mu$, and consider a high-frequency sound wavepacket traveling through such event. If the equations of motion are hyperbolic, we can work in the eikonal approximation\footnote{In a sense, this is equivalent to performing the characteristic manifold analysis mentioned above \cite[\S 1.6]{Rauch_book}} (``fields'' $\propto e^{iS}$, with $S$ large), and we find that the wavepacket locally behaves as a plane wave with dispersion relation \cite{Babichev:2007dw} (in the $S\rightarrow \infty$ limit)
\begin{equation}\label{disperion}
(G^{-1})^{\mu \nu} \, \partial_\mu S \, \partial_\nu S =0 \, .
\end{equation}
Now, let us work in a local reference frame such that $u^\mu=(\gamma,\gamma v,0,0)$, and choose the wavevector to be locally transversal to the background flow, e.g. $\partial_\mu S=(\partial_t S,0,0,\partial_z S)$. With these choices of $u^\mu$ and $\partial_\mu S$, eq. \eqref{disperion}  becomes
\begin{equation}\label{illposed}
\partial_t S =\dfrac{\pm w \, \partial_z S}{\gamma \sqrt{1-v^2 w^2}} \, .
\end{equation}
Now we immediately see the problem. If $vw>1$, the frequency $\partial_t S$ becomes imaginary for real $\partial_z S$. This means that, in the chosen reference frame, there exist local perturbations transversal to the flow that grow exponentially, i.e. $|\text{``fields''}|\sim e^{|\partial_z S|t}$. This is a particularly violent type of instability, because its growth timescale is proportional to $|\partial_z S|^{-1}$, which tends to 0 at large spacelike gradients. It follows that UV modes can grow \textit{arbitrarily} fast. In the mathematics literature, this type of ``infinitely fast instability'' is known as Hadamard ill-posedness \cite[\S 3.10]{Rauch_book}. 

Ill-posed initial-value problems cannot be reliably simulated, because errors due to discretization grow faster as we increase the available wavenumbers (i.e. as we refine the grid)   \cite{Babichev:2007dw,Gavassino:2022nff}. In order to see this schematically, consider a hypothetical sequence of increasingly refined simulations $s_n$, whose grid spacing decays like $\Delta z_n\sim 1/n$, and where the initial numerical rounding error of the field variables decays like $\Delta \Phi_{n}(0,z) \sim 1/n^\alpha$ (with $\alpha >0$). Each simulation $s_n$ is now able to resolve a new Fourier mode with wavevector $k\sim n$, which will be automatically excited by rounding errors with a magnitude of order $1/n^\alpha$. These ``error waves'' will then grow according to the dispersion relation in eq. \eqref{illposed} (assuming $vw>1$), giving
\begin{equation}\label{dphione}
\Delta \Phi_n(t,z) \sim \dfrac{1}{n^\alpha} \sin(nz) \exp\bigg[\dfrac{w nt}{\gamma \sqrt{v^2 w^2-1}} \bigg] \, ,
\end{equation}
We see that, even though, in the limit of large $n$, the initial magnitude of the error wave tends to zero, the same is not true for the magnitude at any later time $t>0$. In fact, the magnitude scales like $\Delta \Phi_n(t,z)\sim e^{nt}/n^\alpha$, which becomes infinitely large for large $n$. In other words, better resolution should lead to worse numerical error.

%%%%%%%%%%%%%%%%%%%%%%%%%%%%%%%%%%%%%%%%%%%%%%%%%%%%%%%%%%%%
%%%%%%%%%%%%%%%%%%%%%%%%%%%%%%%%%%%%%%%%%%%%%%%%%%%%%%%%%%%%
\subsection{Elliptic case}\label{elliptiamo}
%%%%%%%%%%%%%%%%%%%%%%%%%%%%%%%%%%%%%%%%%%%%%%%%%%%%%%%%%%%%
%%%%%%%%%%%%%%%%%%%%%%%%%%%%%%%%%%%%%%%%%%%%%%%%%%%%%%%%%%%%

In Sections \ref{hy1} and \ref{hy2}, we considered instances of acausality where $w^2>1$. However, Israel-Stewart theory can also violate causality in another way. Consider again eq. \eqref{bulkione}, and suppose that $\Pi$ approaches $-(\varepsilon+P)$ from below. Then, $w^2$ becomes negative (i.e. $w$ becomes imaginary). When this happens, the signature of the acoustic metric $G_{\mu \nu}$ becomes $(-,-,-,-)$, and the acoustic cone no longer exists. This implies that the theory is no longer hyperbolic: it becomes elliptic, which results in causality violations of the strongest kind. In fact, using Holmgren's theorem \cite[\S 1.8]{Rauch_book}, one can prove that, in an elliptic theory, if we know the values of the fields in a finite neighborhood of an event, then we can reconstruct the full solution across the whole spacetime, in a way that is completely analogous to the process of analytic continuation (indeed, the Cauchy-Riemann equations are an example of an elliptic system \cite[\S 1.1]{Rauch_book}). Moreover, elliptic equations are known to give rise to ill-posed initial value problems, due to infinitely fast growing waves that are similar in shape to eq. \eqref{dphione}. The only difference is that, in this case, the instability occurs in all reference frames, i.e. for any $v$.

%%%%%%%%%%%%%%%%%%%%%%%%%%%%%%%%%%%%%%%%%%%%%%%%%%%%%%%%%%%%
%%%%%%%%%%%%%%%%%%%%%%%%%%%%%%%%%%%%%%%%%%%%%%%%%%%%%%%%%%%%
%
%
%
%
%%%%%%%%%%%%%%%%%%%%%%%%%%%%%%%%%%%%%%%%%%%%%%%%%%%%%%%%%%%%
%%%%%%%%%%%%%%%%%%%%%%%%%%%%%%%%%%%%%%%%%%%%%%%%%%%%%%%%%%%%
%%%%%%%%%%%%%%%%%%%%%%%%%%%%%%%%%%%%%%%%%%%%%%%%%%%%%%%%%%%%
%%%%%%%%%%%%%%%%%%%%%%%%%%%%

\section{A new instability benchmark test}

\label{sec:simple_test}

As discussed previously, hydrodynamic simulations can enter the acausality-driven unstable regime, at least in some localized regions of space. The question naturally arises as to the manifestation of instabilities in numerical solutions. 
To study this question systematically, it is instructive to have a test scenario where we know analytically how instabilities develop. Before discussing such a solution, we first introduce a nomenclature that helps to summarize the causality-stability status in numerical simulations.

\subsection{Classification of acausalities and instabilities in numerical simulations: the good, the bad, and the ugly}
\label{eq:GBU}
%%%%%%%%%%%%%%%%%%%%%%%%%%%%%%%%%%%%%%%%%%%%%%%%%%%%%%%%%%%%
%%%%%%%%%%%%%%%%%%%%%%%%%%%%%%%%%%%%%%%%%%%%%%%%%%%%%%%%%%%%

In this section's numerical test, we evaluate the characteristic speed $w$ associated with the acoustic cone at every point in the numerical grid used to solve the hydrodynamic equations. If, at a given point, the acoustic cone exists and is contained inside the lightcone, then we know that locally the dynamics is causal, and we label the fluid cell as ``good'' (color-code: green). If instead the acoustic cone falls outside the lightcone, but still points to the future (fig.~\ref{fig:vwcones}, left panel), then we know that the dynamics is locally acausal but stable, and we say that the fluid cell is ``bad'' (color-code: yellow). Finally, if a part of the acoustic cone points towards the past (fig.~\ref{fig:vwcones}, right panel), or if there is no acoustic cone at all\footnote{In general, the non-existence of the acoustic cone is deduced from some characteristic speeds being imaginary, which is analogous to saying that the signature of $G_{\mu\nu}$ is no longer $(-,+,+,+)$, as discussed in Section \ref{elliptiamo}.}, then we know that the dynamics is acausal and unstable, and we say that the fluid cell is ``ugly'' (color-code: red). The existence of red regions in a given snapshot would technically imply that 
the equations would not be reliably solved after that time step (at least in that region), and numerical errors are expected to accumulate due to singularities in the mathematical solution. 

In the case of a fluid with bulk viscosity alone, distinguishing between good, bad, and ugly cells is straightforward:
\begin{itemize}
\item \textbf{Good (i.e.~causal and stable):} We have that $w^2 \in [0,1]$;
\item  \textbf{Bad (i.e.~acausal, but stable):} We have that $w^2>1$, but $vw<1$;
\item \textbf{Ugly (i.e.~unstable):} Either $w^2<0$, or $vw\geq 1$.
\end{itemize}
In the above, the speed $w^2$ is given by eq. \eqref{bulkione}. 

As mentioned earlier, a scenario that would include shear viscosity is more complicated. First of all, the fluid would then be able to send out information both via sound waves and via shear waves (while with bulk alone it can transmit only sound waves). Each of these waves has its own ``characteristic cone'', and the acoustic cone is the one that transmits information with the fastest speed (imaginary speeds count as infinite). This implies that, in practice, all the characteristic speeds must be checked one by one when shear viscosity is present. Additionally, there is the issue that the shear stress introduces local anisotropies in the fluid, and this causes the speeds of the waves to depend on the direction of propagation. Unfortunately, no analytical formula is known for the speed of propagation in a generic direction, except when the signal travels along one of the three principal axes of the stress tensor. Hence, to keep the discussion simple, we shall perform our instability tests in systems with only bulk stress. We note that the classification introduced above is useful not only to the example in the following section, but also for any fluid model with only bulk viscosity.

\subsection{Theory specification}

We consider a bulk viscous fluid at zero chemical potential. The equilibrium equation of state is assumed to be $P=c_s^2 \varepsilon$, where $c_s^2=\text{const} \in (0,1)$ is the speed of sound squared. The stress-energy tensor reads
\begin{equation}\label{Tmn}
T^{\mu \nu} = (bP+\Pi)u^\mu u^\nu +(P+\Pi)g^{\mu \nu} \, ,
\end{equation}
with $b=1{+}c_s^{-2}$. We assume that the bulk stress $\Pi$ obeys an Israel-Stewart-type equation of motion:
\begin{equation}\label{IS}
\tau_\Pi u^\mu \partial_\mu \Pi +\Pi=-\zeta \partial_\mu u^\mu \, ,
\end{equation}
where the relaxation time $\tau_\Pi$ is a constant, and $\zeta=aP\tau_\Pi$ ($a>0$ is another constant). Using eq. \eqref{bulkione}, we find that, in the non-linear regime, the speed of propagation of information is
\begin{equation}\label{w2}
w^2 = c_s^2 + \dfrac{a}{1+c_s^{-2}+\Pi/P} \, ,
\end{equation}
which we require to be causal at least near equilibrium (i.e. for $\Pi=0$). This gives the inequality $a\leq (1-c_s^2)(1+c_s^{-2})$. Still, causality is violated when $\Pi$ is sufficiently large and negative, specifically when
\begin{equation}
\dfrac{\Pi}{P} < -1 -c_s^{-2}+ \dfrac{a}{1-c^2_s} \, .
\end{equation}
We also note that, in the present scenario, because of the functional form of the characteristic speed, ellipticity (i.e.~$w^{2}<0$), only occurs for an interval of even more negative $\Pi$, namely $-(1+c_{s}^{-2}(1+a)) < \Pi/P < - 1 - c_{s}^{-2}$. When taking $\Pi$ to very negative values, this is only accessible after $w^{2}$ diverges at $\Pi/P = -1-c_{s}^{2}$.

\subsection{Flow geometry}

Choosing a preferred reference frame, we consider a homogeneous slab of fluid that travels in the positive $x^1$ direction. As usual, ``homogeneity'' means that all quantities are assumed to depend only on $t$ (in the chosen reference frame). Hence, the conservation laws $\partial_\mu T^{\mu \nu}=\partial_t T^{0\nu}=0$ imply that the energy density $T^{00}$ and the momentum density $T^{01}$ are constant in time. In a non-relativistic context, constancy of $T^{01}$ would imply that the speed of motion $v(t)$ is also constant in time, because velocity and momentum are the same up to a constant. However, in relativity, the situation is different, because the momentum depends also on $\Pi$, which is an independent degree of freedom. As we shall show below, the fluid can use the bulk stress $\Pi(t)$ as ``rocket fuel'' to power its own acceleration.

First of all, let us use spatial homogeneity to simplify the equations of motion. Our degrees of freedom are the functions $\{P(t),v(t),\Pi(t) \}$, where $v=u^1/u^0$ is positive (by assumption\footnote{The assumption  $v > 0$ is not strictly required in the derivation of eq.~\eqref{garzino}. However, 
in practice, after solving the equation of momentum conservation, one finds that the sign of $v$ cannot change, and so it may be taken positive.}). The equations of motion are the following:
\begin{equation}\label{garzino}
\begin{split}
& \dfrac{d}{dt}\big[(bP{+}\Pi)\gamma^2 -P-\Pi\big]=0 \, , \\
& \dfrac{d}{dt}\big[(bP{+}\Pi)\gamma^2 v\big]=0 \, , \\
& \tau_\Pi \gamma \dfrac{d\Pi}{dt} +\Pi=-\zeta \dfrac{d\gamma}{dt}  \, , \\
\end{split}
\end{equation}
where $\gamma=(1{-}v^2)^{-1/2}=u^0$ is the usual Lorentz factor. The first and the second equations are the conservations of energy and (the first component of) momentum, see eq.~\eqref{Tmn}. The third is just the Israel-Stewart equation of motion of the bulk stress, see eq. \eqref{IS}. We can immediately integrate the first two, denoting the two integration constants by respectively $\mathcal{E}$ (which is just $T^{00}$) and $\mathcal{P}^1$ (which is just $T^{01}$), so that\footnote{Since the fluid is assumed to travel towards positive $x^1$, we have that $0<\mathcal{P}^1<\mathcal{E}$, where the second inequality is the dominant energy condition \cite{Hawking:1973uf}.}
\begin{equation}\label{pppppp3545}
\begin{split}
& (bP{+}\Pi)\gamma^2 -P-\Pi=\mathcal{E} \, , \\
& (bP{+}\Pi)\gamma^2 v =\mathcal{P}^1 \, . \\
\end{split}
\end{equation}
This is an algebraic system, which we can solve for $P$ and $\Pi$. The result is
\begin{equation}\label{garzone}
 \begin{split}
P={}& c_s^2 (\mathcal{E}{-}\mathcal{P}^1 v) \, , \\
\Pi={}& \bigg[ \dfrac{1}{v}+ c_s^2 v -(1{+}c_s^2)\dfrac{\mathcal{E}}{\mathcal{P}^1} \bigg] \mathcal{P}^1  \, . \\
\end{split}
\end{equation}
It follows that, once the integration constants $\mathcal{E}$ and $\mathcal{P}^1$ are fixed, all physical quantities (including $P$, $\Pi$, $\zeta$, $\tau_\Pi$, and $w$) can be written as functions of $v$ alone. Therefore, the third equation of \eqref{garzino} can be interpreted as an ordinary differential equation for $v(t)$, which can be recast (after some algebraic manipulations) into the following form:
\begin{equation}\label{taugamone}
\tau_\Pi \gamma(v) \dfrac{dv}{dt} = \dfrac{v^2 \Pi(v)/\mathcal{P}^1}{1-v^2 w(v)^2} \, .
\end{equation}

\subsection{Mathematical considerations}

We recall that, according to the discussion in Section \ref{hy1}, an instability should show up when $vw$ grows above 1. Indeed, this is precisely what we see in eq.~\eqref{taugamone}. Specifically, we note that, when $vw \equiv 1$, the right-hand side of eq.~\eqref{taugamone} is divergent, which is a clear indication that the initial value problem has become illposed. Moreover, it is straightforward to show that, if we approach $vw=1$ from below (i.e. if we come from the bad region), then $dv/dt$ diverges to $-\infty$, whereas if we approach $vw=1$ from above (i.e. if we come from the ugly region), then $dv/dt$ diverges to $+\infty$. In other words, the state with $vw=1$ is a bifurcation point, where two distinct solutions are possible, one where the fluid decelerates towards equilibrium (in the bad case), and one where the fluid accelerates away from equilibrium (in the ugly case).\footnote{Here, the ``equilibrium state'' is just the state with $\Pi=0$. To see that $dv/dt\,{>}\,0$ points away from equilibrium in a neighbourhood of $vw=1$, we can just note that, since we are in the acausal region, $\Pi$ is negative, which implies that $v$ is larger than the equilibrium value, by the second equation of \eqref{pppppp3545}. So, if we wish to approach equilibrium, we would need $v$ to decrease.} Once again, this confirms that the ugly states experience reversed dissipation, as they tend to move away from equilibrium, rather than towards it.

\subsection{An explicit solution}

\begin{figure}[h!]
    \centering
\includegraphics[width=0.9\linewidth]{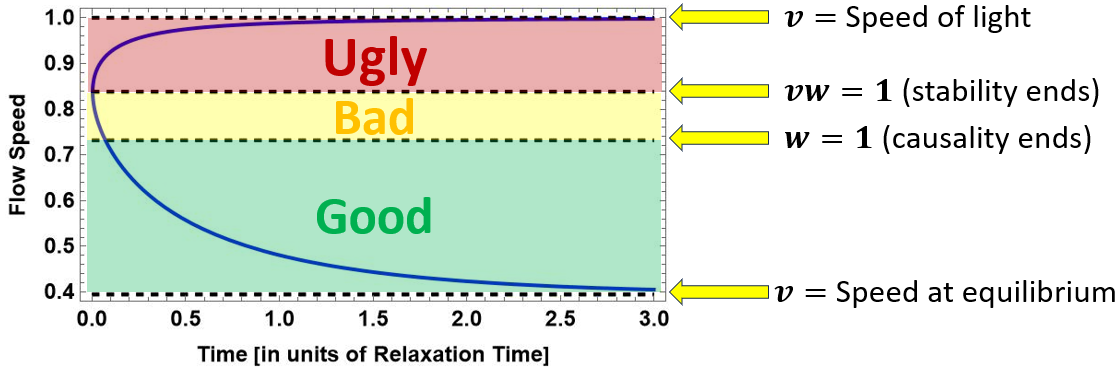}
\caption{Graph of the function $v(t)$ obtained by inverting eq.~\eqref{elboss}. This is an analytical solution of Israel-Stewart theory with only bulk. Around $v\sim 0.732$, the system exits the causal region (good), and enters the acausal but stable region (bad). Nothing serious happens. However, around $v_c \sim 0.838$, the fluid enters the acausal and unstable region (ugly), and the solution bifurcates. The new solution branch is pathological, since $\Pi$ keeps getting more and more negative, so the fluid accelerates to the speed of light. Note that, when $v\rightarrow 1$, the ratio $\Pi/P$ saturates to $-4$, which is a finite value.}
    \label{fig:Instabone}
\end{figure}

We can solve eq.~\eqref{taugamone} analytically. For definiteness, we choose the parameter values: $\{c_s^2,a,\tau_\Pi,\mathcal{E}/\mathcal{P}^1 \}=\{1/3,1,1 \textrm{(a.u.)},2 \}$. Then, eq.~\eqref{taugamone} becomes
\begin{equation}\label{dvdtheta}
\dfrac{dv}{dt} = \dfrac{v}{\gamma(v)^3} \dfrac{v^2-8v+3}{2v^4-2v^3-4v^2+3} \, ,
\end{equation}
which produces the following analytical solution:
\begin{equation}\label{elboss}
\begin{split}
t(v)={}& t_0+ \frac{1}{312} \Bigg[ 78 \gamma(v) -\sqrt{39 \left(1586 \sqrt{13}-5155\right)} \arctan\left(\frac{\sqrt{\frac{1}{2} \left(\sqrt{13}-1\right)} }{(1+v)\gamma(v)}\right) \\
& -624 \, \text{arctanh} \left(\frac{1}{\gamma(v)(1+v)}\right)+\sqrt{39 \left(1586 \sqrt{13}+5155\right)} \text{arctanh}\left(\frac{\sqrt{\frac{1}{2} \left(\sqrt{13}+1\right)} }{(1+v)\gamma(v)}\right) \Bigg] \, , \\
\end{split}
\end{equation}
see fig. \ref{fig:Instabone}. Given the choice of parameters above, the bifurcation between bad and ugly solutions occurs at the critical velocity $v_c\approx 0.838$, where $vw=1$. Above this velocity, the fluid spontaneously accelerates indefinitely, and its speed gets closer and closer to the speed of light (the Lorentz factor grows approximately as $\gamma \sim t$).
    
It is also verified that causality violation by itself does not imply instability. In fact, in the present example, the characteristic speed $w$ becomes greater than 1 for $v>\sqrt{3}{-}1 \approx 0.732$. Yet, in the ``bad'' region (where $\sqrt{3}{-}1<v<v_c$), the dynamics remains stable. To see the instability, we need the acausal characteristic to have inverted chronology, which happens only in the ugly region.

\subsection{Numerical simulations}
\label{sec:num-sol-bulk}

\begin{figure}[t]
	\centering
		\includegraphics[scale=0.4]{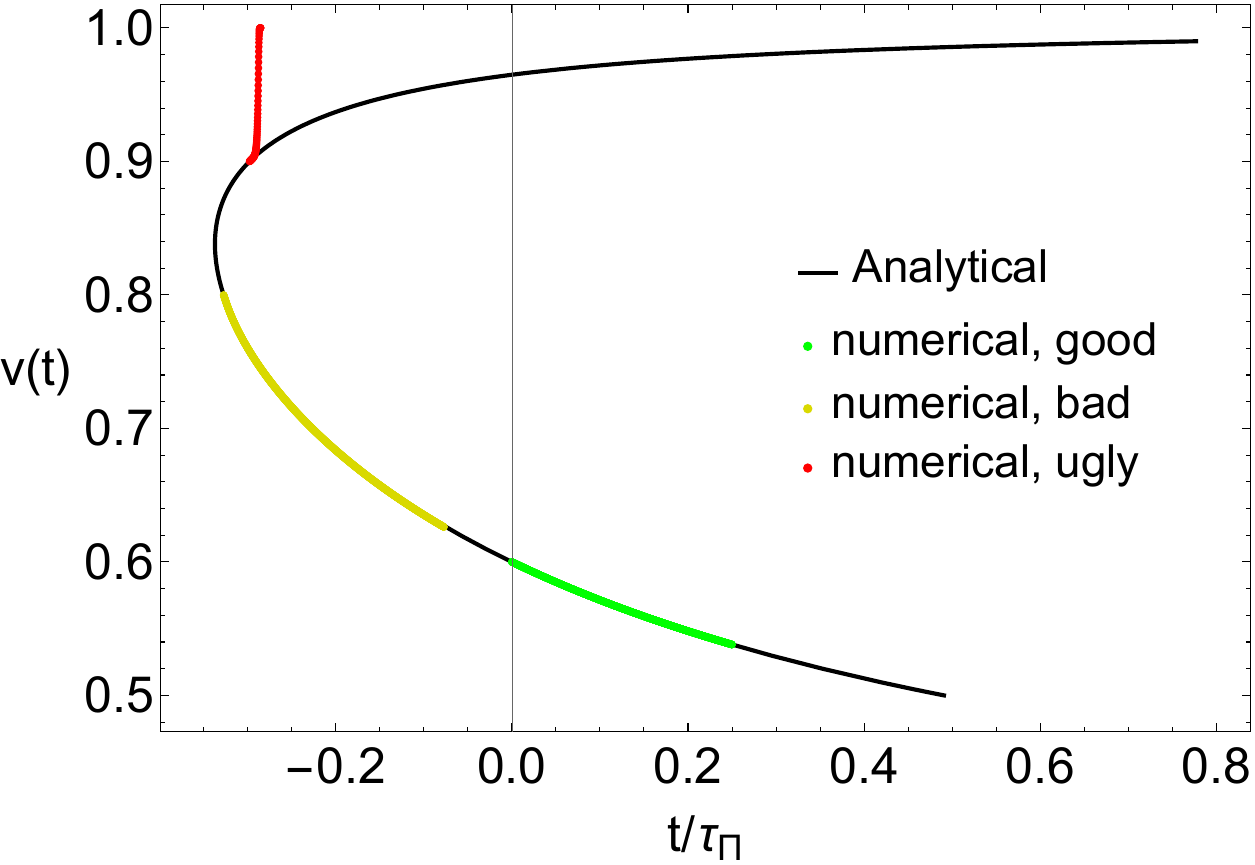}
	\includegraphics[scale=0.4]{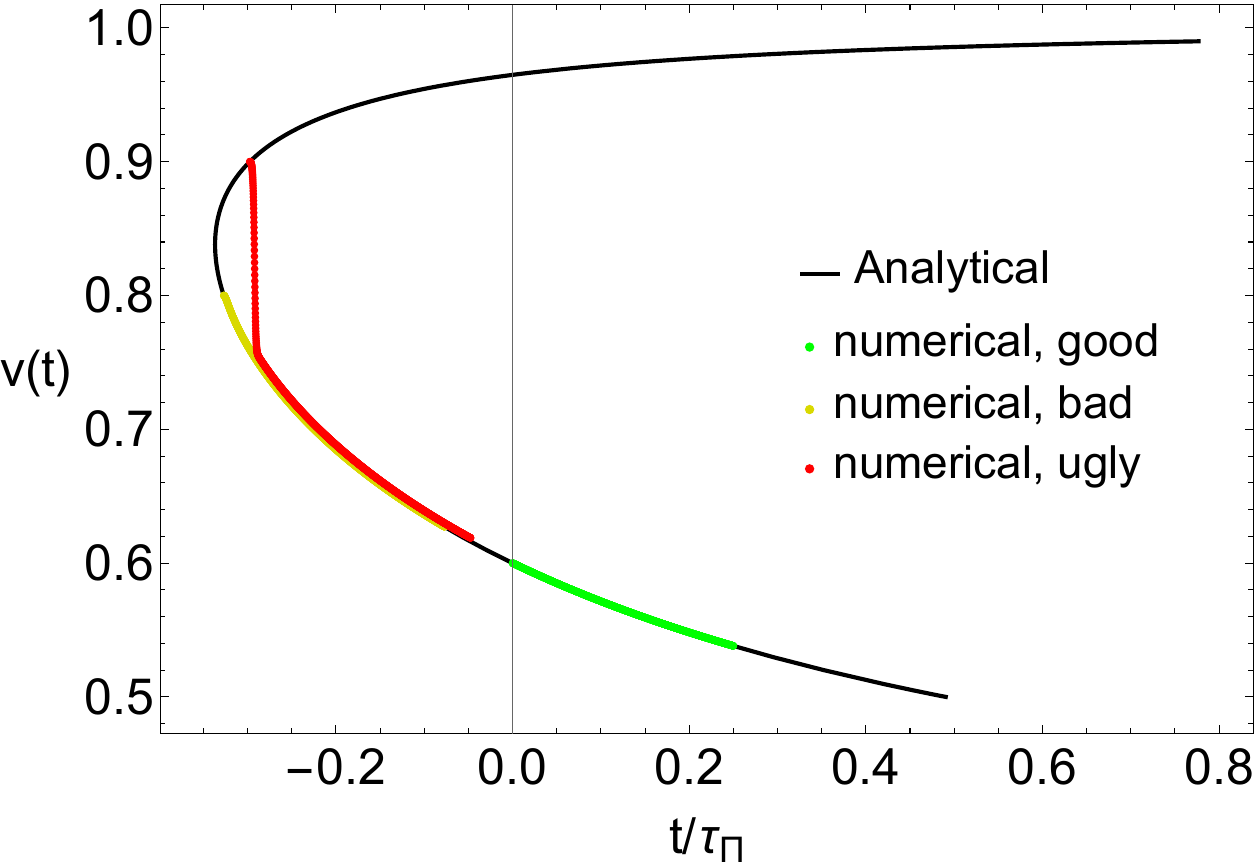}
	\caption{Comparison between the analytical solution in eq.~\eqref{elboss} and numerical solutions with initial conditions with different causality/stability status. (Left Panel) Exact initial condition.  (Right Panel)  ``Zero-time-gradient'' initial conditions. See text for details.
	}
	\label{fig:GBU-preIC}
\end{figure}

In order to assess the numerical effect of causality violations and instabilities in a controlled manner, we perform simulations on the system described in the previous subsection and employ the numerical solver MUSIC \cite{Schenke:2010nt,Schenke:2010rr,Paquet:2015lta}. The MUSIC solver uses hyperbolic $(\tau,x,y,\eta_s)$ coordinates with $t=\tau \cosh \eta_s$ and $z=\tau \sinh \eta_s$ rather than Cartesian $(t,x,y,z)$ coordinates. We sidestep this  by taking the initial hyperbolic time to be very large, namely, $\tau = 10^{6}$ fm/c, such that $\tau \approx t$. Periodic boundary conditions are employed in the longitudinal ($z$) direction, which is aligned with the fluid velocity. Throughout this section, we employ the parameter set $\{c_s^2,a,\tau_\Pi,\mathcal{E}/\mathcal{P}^1 \}=\{1/3,1,1 \text{ fm},2 \}$. We compare the behavior of the numerical solution with respect to the analytical solutions in eqs.~\eqref{garzone} and \eqref{elboss}.  In the left panel of fig.~\ref{fig:GBU-preIC}, we display numerical solutions for $v(t)$ for `good' initial conditions ($v(0) = 0.6$; causal), `bad' initial conditions ($v(0) = 0.8$; acausal but stable), and `ugly' initial conditions ($v(0) = 0.9$, where $vw > 1$). The analytical solution is displayed as the thinner black line.
	The right panel of fig.~\ref{fig:GBU-preIC} corresponds to a different numerical solution that we discuss in the next paragraph.
	In the `good' and the `bad' regimes, we readily see
	that the numerical solutions accurately reproduce the expected analytical results. In the unstable (`ugly') regime, a numerical runaway solution which completely violates the analytical solution is obtained, as expected from the discussion in Section~\ref{hy2}. At $t/\tau_{\Pi} \sim 0.015$, the runaway numerical solution approaches $v(t) \approx 1$. From this point, we expect the numerical solution to be dependent on implementation choices in the solver designed to handle problematic solutions\footnote{
	There is a regulator in MUSIC acting in regions of very low energy density that  artificially reduces the bulk stress (and also shear stresses, when present) when their ratio to the energy density plus pressure (a proxy for the inverse Reynolds number) is too large.
	Regulators of this sort are common in other viscous hydrodynamic solvers, at least for Eulerian solvers.
	In simulations of heavy-ion collisions, this regulation is typically acting in low-density regions where fluid dynamics is not expected to hold, but where the hydrodynamics equations are still solved due to the set-up of the problem. This regulator is disabled for all results shown in this work.
	We did verify that turning on this regulation does not affect the solutions in this section with `good' or `bad' initial conditions. However, for ugly initial conditions, the code with the regulator enabled performs another `jump' in $v(t)$ such that $v(t) \approx 0$ for the remainder of the evolution after $v(t) \approx 1$. In this regime, even the error estimators discussed in Appendix~\ref{sec:hom-fluid-obed} grow by orders of magnitude. An analogous behavior is observed for eq.~\eqref{garzone} for $P(v)$ and $\Pi(v)$, which are assessed also in Appendix \ref{sec:hom-fluid-obed}.
	\label{fn:QRV-MUSIC}
    }.
	 Thus, we conclude that the conservation laws are consistently solved numerically for `good' and the `bad' initial conditions and violations are seen in the `ugly' regime.
		
	There is, however, an issue with obtaining the results from the left panel of fig.~\ref{fig:GBU-preIC}; it is related to an implementation choice of the relativistic hydrodynamic solver MUSIC and of other  widely-employed solvers used for heavy-ion collision simulations \cite{Paquet:2015lta,Shen:2014vra,Nijs:2020roc,Molnar:2009tx}. 
	 In general, the local conservation laws, $\partial_{\mu}T^{\mu \nu} = 0$, and the bulk relaxation equation $\tau_\Pi u^\mu \partial_\mu\Pi + \Pi = -\zeta\partial_\mu u^\mu$ are employed to solve an initial value problem where the fields $(\varepsilon, u^{\mu}, \Pi)$ are given at some initial time $t_0$. However, for the ease of numerical implementation, these equations are often rewritten as
	\begin{align}
		\partial_t T^{t\mu}_{\rm id} = -\partial_i T^{i\mu}_{\rm id} - \partial_\nu T_{\rm diss}^{\nu\mu}\label{eq:idealevolution-mt},\\
		\partial_t\Pi = -\frac{1}{\gamma\tau_\Pi}\left(\Pi + \zeta\partial_{\mu}u^{\mu}\right),\label{eq:bulkevolution-mt}
	\end{align}
	where the decomposition $T^{\mu\nu} = T^{\mu\nu}_{\rm id} + T_{\rm diss}^{\mu\nu}$, was made, $T^{\mu\nu}_{\rm id} = \varepsilon u^{\mu}u^{\nu} + P(\varepsilon)(g^{\mu \nu} +  u^\mu u^\nu)$ is the ideal component of the energy-momentum tensor and, since bulk viscosity is the only source of dissipation in the present case, $T_{\rm diss}^{\mu\nu} = \Pi(g^{\mu\nu} + u^\mu u^\nu)$ is the dissipative component of the energy-momentum tensor. The right-hand sides of the above equations are treated as ``source terms'' that are evaluated first at each time step. The value of the hydrodynamic fields at following time steps can then be calculated.
	A challenge of this approach is that the right-hand ``source terms'' still have temporal derivatives. Therefore, it is necessary to know $\partial_t T_{\rm diss}^{0\mu}$ in eq.~\eqref{eq:idealevolution-mt} and $\partial_t\gamma$ in eq.~\eqref{eq:bulkevolution-mt} to get the source terms. This is done by storing the field values from the previous time step. Changes in the fields between the previous and current time steps are used to evaluate the temporal derivatives in the source terms.  This approach works for all time-steps except the first one. This shortcoming is often handled by effectively setting the temporal derivatives to zero at this first time-step. This is generally numerically implemented by assuming that the state of the system, given by $(\varepsilon, u^{\mu}, \Pi)$ at time $t_0$, and the state at a fictitious time $t_0-\delta t$ are identical. 
	In what follows, we refer to this implementation choice as `zero-time-gradient initialization'.
	This zero-time-gradient initialization can lead to violations of the equations of motion and it is not clear to what extent this should impact generic solutions.  
	For the simple example studied in this section, the zero-time-gradient initialization can be circumvented by initializing properly the gradients in the initial time step with the analytical values from eqs.~\eqref{pppppp3545} and \eqref{dvdtheta}. The result is what is displayed in the left panel of fig.~\ref{fig:GBU-preIC}. 
	In Appendix \ref{apn:details-IC}, we outline a numerical iterative procedure, valid in generic flow configurations, to compute the gradients at initial time while still using the source term approach. 
	
	In the context of our benchmark solution, by using a small timestep ($\delta t = 0.0001 \tau_{\Pi}$), we recover the analytical results in the `good' and the `bad' regimes even with the first timestep being solved incorrectly due to the zero-time-gradient initialization, as shown in the right panel of fig.~\ref{fig:GBU-preIC}. 
	In the `ugly' regime, however, the zero-time-gradient initialization leads to more drastic consequences: the system quickly `jumps' from the unstable to the stable branch of the curve for $v(t)$, which is not the correct solution given the initial conditions. It thus appears that the zero-time-gradient initialization hides the effect of the instability in the numerical solution, at least for this specific example.

\section{Causality-stability test on a two-dimensional bulk-only fluid}
\label{eq:GBU-test-2D}

We now illustrate how acausal and unstable regions can emerge from initial conditions used to study relativistic heavy-ion collisions.
We consider a longitudinally boost-invariant hydrodynamic system with only bulk stresses, so that the energy-momentum tensor is given by
\begin{equation}
\label{eq:ener-mom-bulk}
    T^{\mu \nu} = \varepsilon u^\mu u^\nu +(P(\varepsilon)+\Pi)(g^{\mu \nu} +  u^\mu u^\nu),
\end{equation}
where the functional form of the pressure in terms of the energy density, $P(\varepsilon)$, is provided by the lattice QCD equation of state \cite{HotQCD:2014kol,Bernhard:2018hnz}. 
The local conservation laws for energy and momentum, $\partial_{\mu}T^{\mu \nu} = 0$, are coupled to a minimal Israel-Stewart model, 
\begin{align}
	\label{eq:min_IS-model}
	\tau_\Pi u^\mu \partial_\mu\Pi + \Pi &= -\zeta\partial_\mu u^\mu.
\end{align}   
For the bulk viscosity, we use the parametrization~\cite{Schenke:2020mbo}
\begin{align*}
\frac{\zeta}{s}(T) = A \exp\left[-\frac{(T - T_{\rm peak})^{2}}{\sigma(T)^{2}}\right]
\end{align*}
with $T_{\rm peak} = 0.160$ GeV,  $A = 0.13$, $\sigma(T < T_{\rm peak}) = 0.01$, $\sigma(T > T_{\rm peak}) = 0.12$, and $s = (\varepsilon + P)/T$ is the entropy density. The relaxation time parametrization is from ref.~\cite{Denicol:2014vaa}
\begin{align*}
\tau_{\Pi} = \frac{1}{15}\frac{\zeta}{(\varepsilon + P)(1/3 - c_{s}^{2})^{2}}.  
\end{align*}
We initialize the energy-momentum tensor in eq.~\eqref{eq:ener-mom-bulk} with the IP-Glasma model \cite{Schenke:2012wb,Schenke:2012hg}.\footnote{The input file employed is publicly available at \url{https://github.com/MUSIC-fluid/MUSIC/blob/public_stable/example_inputfiles/IPGlasma_2D/input/epsilon-u-Hydro-t0.6-0.dat}.} This setup is a subset of a widely employed hydrodynamic framework in heavy-ion collision phenomenology, which usually includes also shear-stress effects, second-order transport coefficients and couplings between the different stresses~\cite{Romatschke:2017ejr,Jeon:2015dfa,DerradideSouza:2015kpt,Gale:2013da}.

In connection with the discussion about the zero-time-gradient initialization (see discussion around eqs.~\eqref{eq:idealevolution-mt} and \eqref{eq:bulkevolution-mt}), we note that the impact of neglecting temporal derivatives at the first time step may be smaller in heavy-ion collision simulations. These simulations generally use Milne coordinates where the longitudinal expansion is built in the coordinate system. Milne coordinates have hyperbolic time $\tau = \sqrt{t^2-z^2}$ and longitudinal rapidity $\eta = \tanh^{-1}(z/t)$. The metric tensor is given by $g^{\mu\nu} = {\rm diag.}(-1, 1, 1, 1/\tau^2)$. Equations are initialized at small hyperbolic time $\tau_i$. The expansion rate, in Milne coordinates, is
\begin{equation}
    \nabla_\mu u^\mu = \frac{1}{\sqrt{-g}} \partial_{\mu}(\sqrt{-g} u^{\mu}) = \frac{u^\tau}{\tau} + \partial_\tau u^\tau + \partial_xu^x + \partial_yu^y + \partial_\eta u^\eta,
\end{equation}
where $\nabla_\mu$ denotes the covariant derivative, since curvilinear coordinates are being employed. As we see, the expansion rate is partially accounted for by the $u^\tau/\tau$ term, which stems from the Christoffel symbol. Setting the initial temporal derivative to zero amounts to $\partial_\tau u^\tau = 0$. If $\tau_i$ is small and $u^\tau/\tau_i \gg \partial_\tau u^\tau$, then setting $\partial_\tau u^\tau = 0$ may not be a significant effect in practical terms.

\begin{figure}[t]
\centering
\includegraphics[scale=0.36]{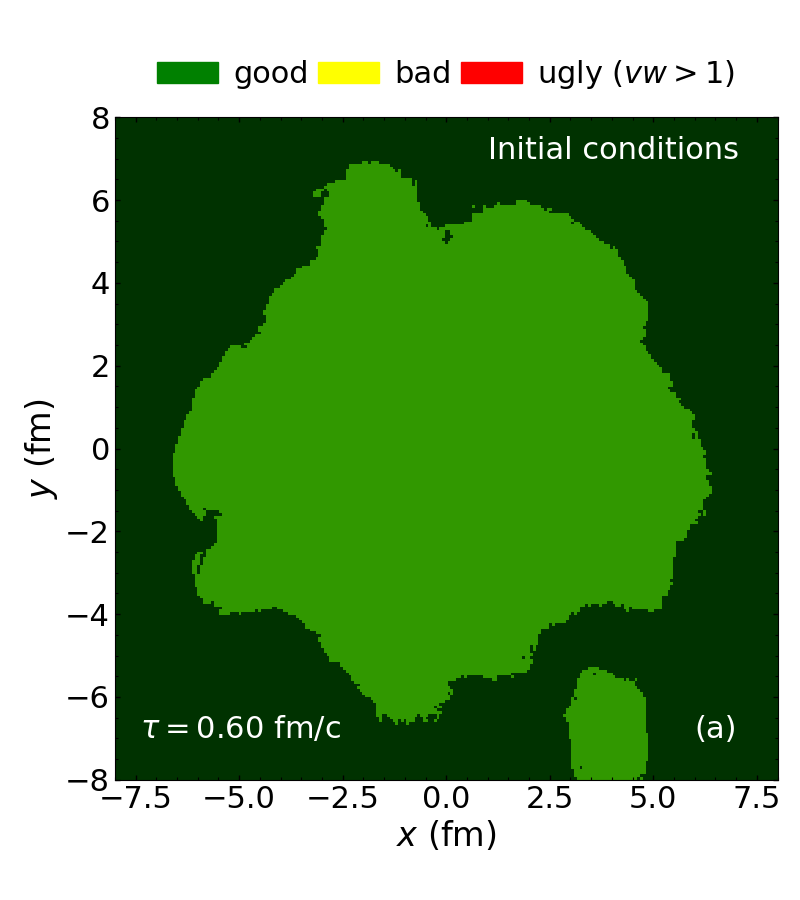}
    \includegraphics[scale=0.0875]{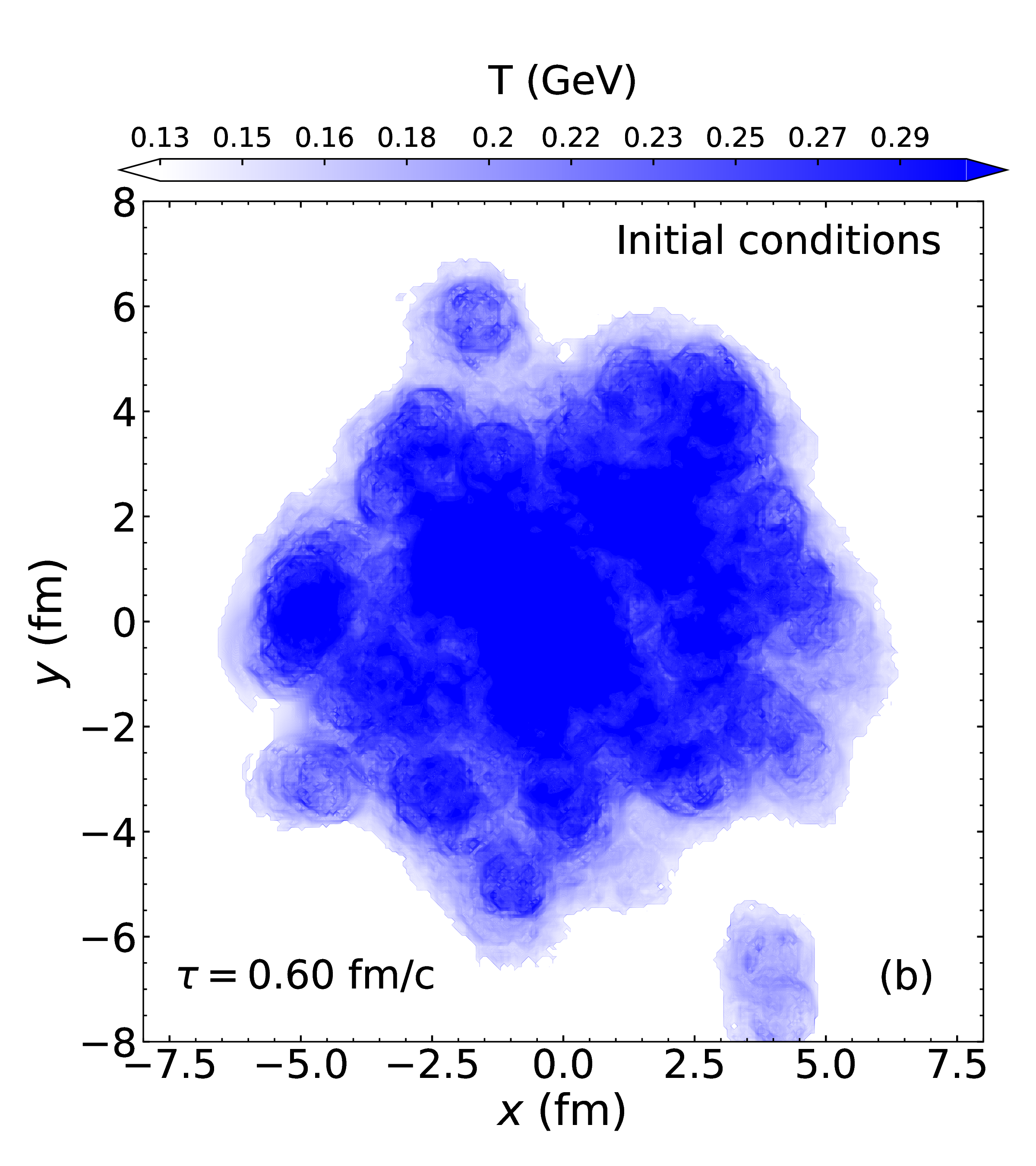}
\includegraphics[scale=0.36]{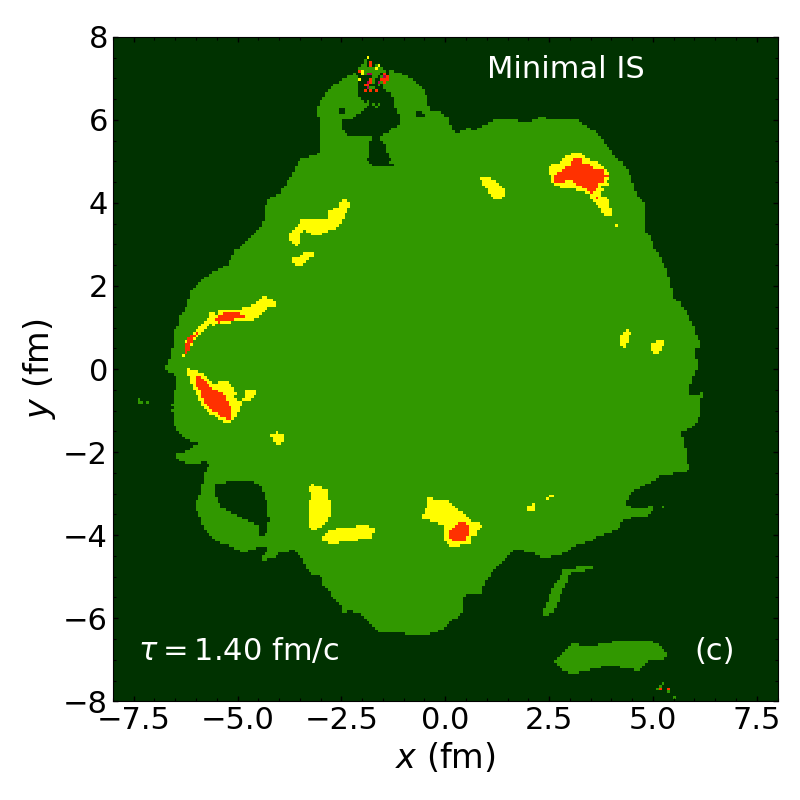}
\includegraphics[scale=0.085]{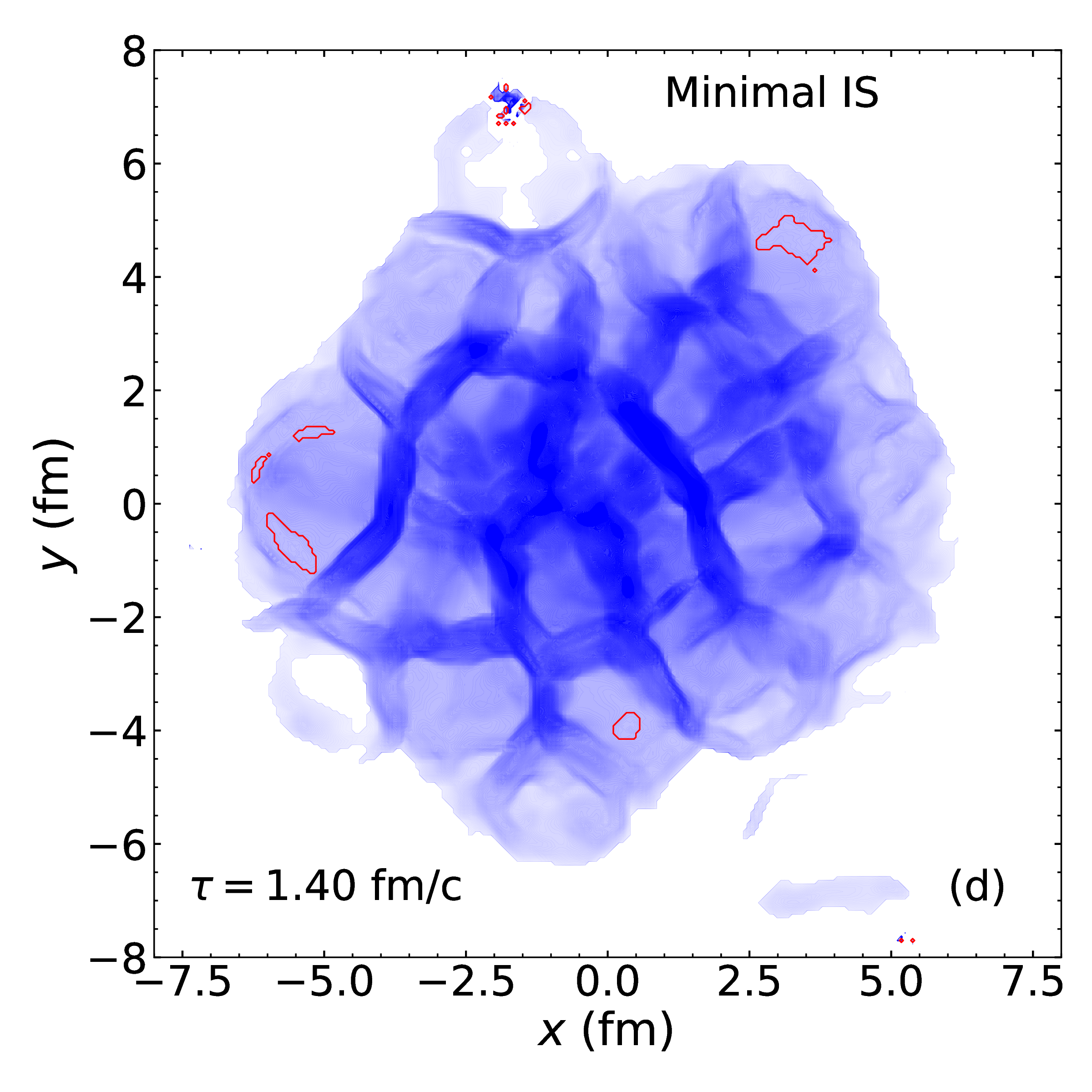}
\includegraphics[scale=0.36]{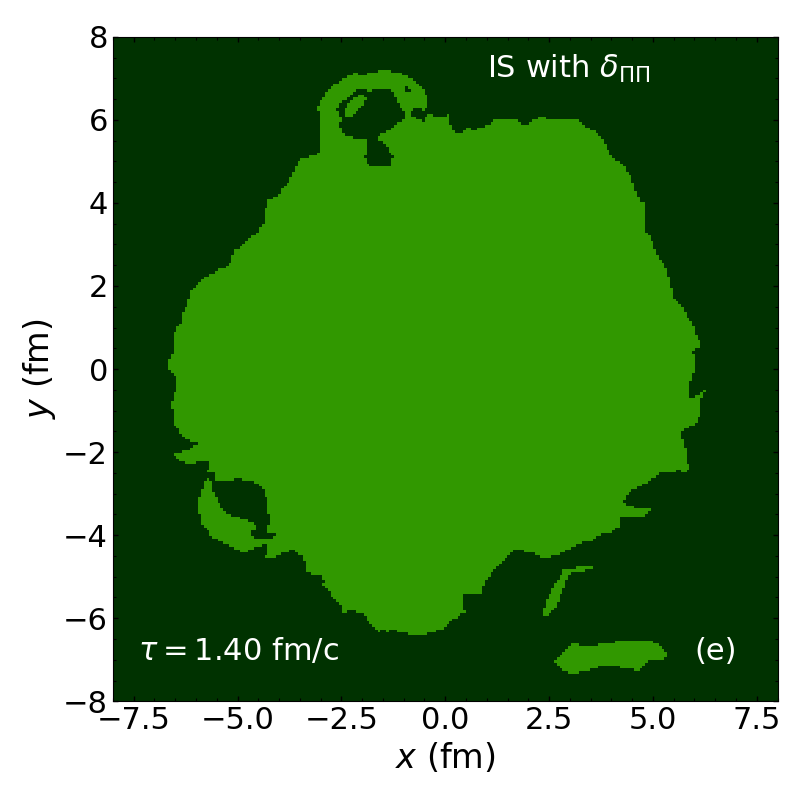}
\includegraphics[scale=0.085]{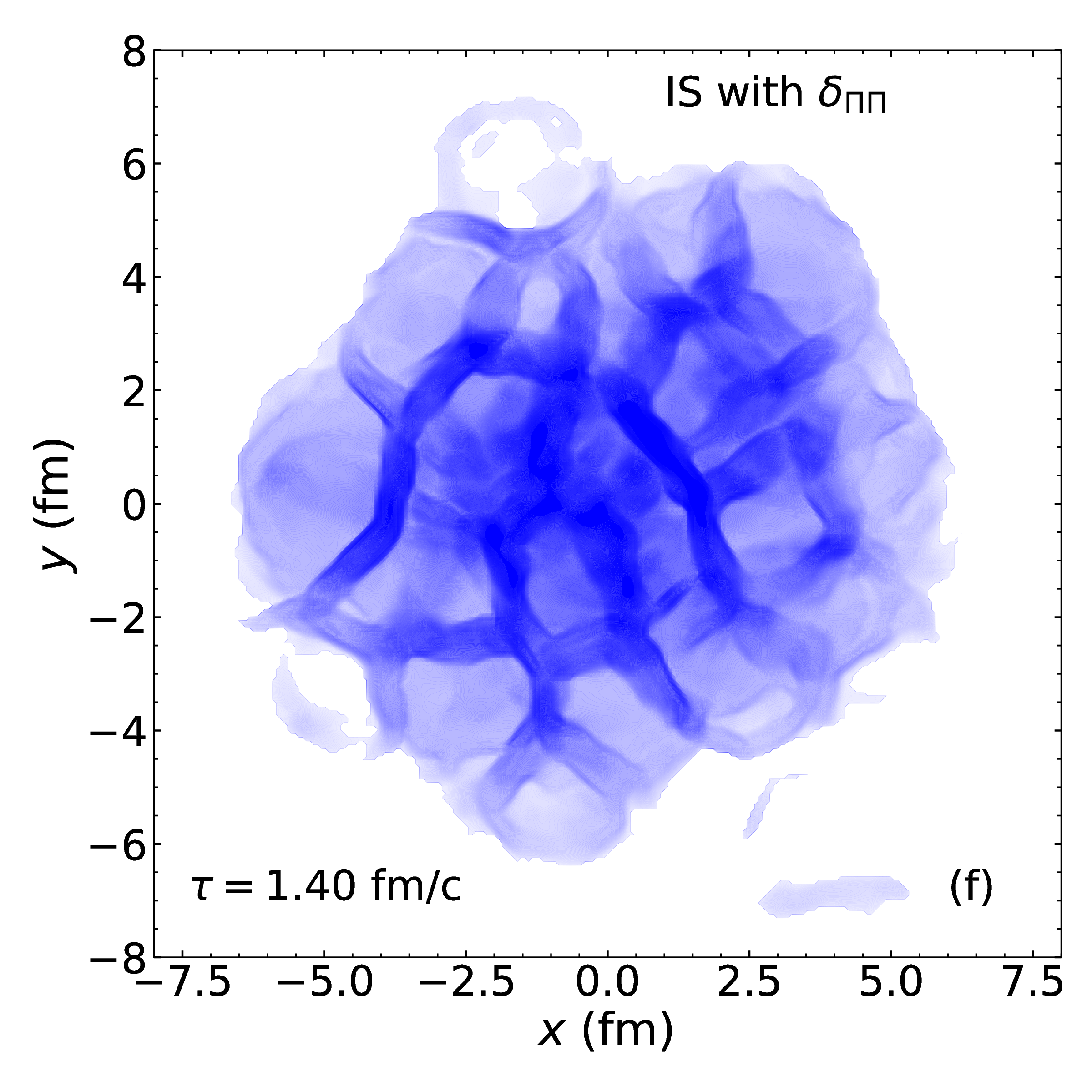}
\vspace{-0.5cm}
\caption{Causality-stability status (panels (a), (c) and (e)) and temperature field ((b), (d) and (f) panels) along the hydrodynamic evolution for the initial (panels (a) and (b)) and later ((c) and (d); (e) and (f)) times for a pure-bulk fluid. In the temperature field plots, the red lines represent the boundaries of the ugly regions of the $vw > 1$ kind. A minimal Israel-Stewart model (`Minimal IS', eq.~\eqref{eq:min_IS-model}) is employed in the center panels and the Israel-Stewart model with a second order transport coefficient (`IS with $\delta_{\Pi \Pi}$', eq.~\eqref{eq:w2-2nd-IS}) in the lower panels.}
\label{fig:GBU-min_IS-2D}
\end{figure}

The hydrodynamic evolution is assessed in fig.~\ref{fig:GBU-min_IS-2D}, which is also produced with MUSIC.\footnote{We note that the regulator mentioned in Footnote \ref{fn:QRV-MUSIC} was also deactivated for the simulations in the present section.} 
Panels (a) and (b) correspond to the initial conditions at $\tau = 0.6$ fm/c, and panels (c) and (d) display the results of the hydrodynamic evolution after $0.8$~fm/c of evolution ($\tau=1.4$~fm/c). The left panels show the `good', `bad', `ugly' color scheme, while right panels show the temperature profile. In panel (a), we see that the characteristic speed is real and subluminal everywhere and thus the system is `globally good' at the initial time. In panel (b), we see a  temperature profile containing sharp substructures that are typical of the initial state model employed. Nevertheless, at $\tau \approx 0.90$ fm/c (not displayed in the figure), acausal regions start to emerge. This contrasts with a possible intuition that one may draw from e.g.~\cite{Plumberg:2021bme}, where acausal regions are largest at initial times and then they shrink, or from the previous section, where the system started in acausal/unstable regions. Thus, the emergence of acausality in the evolution cannot necessarily easily be assessed from the initial conditions. After that, at $\tau \approx 1.10$ fm/c (also not displayed in the figures), ugly regions with $v w \geq 1$ appear. 

In panels (c) and (d) of fig.~\ref{fig:GBU-min_IS-2D}, we see a time frame at $\tau = 1.40$ fm/c following eq.~\eqref{eq:min_IS-model} for $\Pi$. We note, in panel (c), that ugly ($vw>1$) regions are contained within the bad regions. This is expected given the fact that the product $vw$ can only exceed $1$ if $w > 1$, since $v$ is assumed to be subluminal ($v < 1$). From panel (d), we also remark that the acausal regions emerge near the boundary of some ``local wavefronts'' where the expansion is violent enough to drive $\Pi/(\varepsilon + P)$ to sufficiently large and negative values (taking into account, of course, the equation of state and the specific functional form of the transport coefficients). 

We also note that ugly regions with $w^{2}<0$, which do not emerge in the simulations displayed, would have to be contained within the ugly ($vw>1$) regions. This can be understood from the functional form of $w^{2}$, see eq.~\eqref{bulkione}. There, we can see that, by decreasing $\Pi/(\varepsilon + P)$ within negative values, $w^{2}$ can only become negative after $w^{2}$ becomes very large, formally infinite, at $\Pi/(\varepsilon + P) \to -1$. This is specific to the bulk-only case: if shear viscosity is included, $w^{2}<0$ regions might emerge without necessarily going through the bad or the ugly ($vw>1$) regime.

Acausalities and instabilities may change for better or worse when other effects are considered, such as second-order transport coefficients or shear stresses. This stems from the fact that, in general, both the solution of the equation of motion and the formula(s) for the characteristic speed(s) will change. As an example of the effect of second-order transport processes, let us consider changing the minimal Israel-Stewart, eq.~\eqref{eq:min_IS-model}, to 
\begin{equation}
\label{eq:w2-2nd-IS}
\tau_\Pi u^\mu \partial_\mu\Pi + \Pi = -\zeta\partial_\mu u^\mu - \delta_{\Pi \Pi} \Pi \partial_\mu u^\mu  
\end{equation}
where $\delta_{\Pi \Pi}/\tau_{\Pi} = 2/3$ \cite{Denicol:2014vaa}, and leaving all other model parameters unaltered. In this case, the speed of propagation of information $w$ is
\begin{equation}
w^{2} = c_s^2 +\dfrac{\zeta + \delta_{\Pi \Pi} \Pi}{\tau_\Pi (\varepsilon{+}P{+}\Pi)}.   
\label{eq:w2_dpipi}
\end{equation}
With the same initial conditions and evolving for the same amount of time as in the first part of this section, equations of motion \eqref{eq:w2-2nd-IS} lead to an evolution that is entirely causal and is fully stable. This is shown in panels (e) and (f) of fig.~\ref{fig:GBU-min_IS-2D} at time $\tau = 1.40$ fm/c. Since the system was initialized with the same input as the minimal Israel-Stewart model, the initial temperature profile is the same, and we verified that the initial causality status is also the same even though the values of the characteristic speed change (eq.~\eqref{eq:w2_dpipi} instead of eq.~\eqref{bulkione}). We also remark that even though the minimal Israel-Stewart and the model of eq.~\eqref{eq:w2-2nd-IS} have different causality status, the evolution of the temperature is very similar as one may attest by comparing panels (d) and (f) of fig.~\ref{fig:GBU-min_IS-2D}. 

When shear stresses are considered, the situation is expected to become more involved because of the multiple characteristic velocities and local anisotropies already mentioned. Moreover, the coupling between shear and bulk renders the development of intuition quite an intricate task. Using realistic simulations of heavy-ion collisions, ref.~\cite{Chiu:2021muk} found a strong dependence of causality conditions on second-order transport coefficients; within their model, two of the transport coefficients that they found to have the most impact on causality were in fact shear-bulk couplings.

%For instance, in ref.~\cite{Chiu:2021muk}, it was found that in order to satisfy causality in realistic simulations of heavy-ion collisions, three second-order transport coefficients (with a specific parametrization stemming from kinetic theory) are responsible for strong restrictions in the range of values of $\Pi$ and $\pi^{\mu \nu}$. Nevertheless, they also find that the strength of the restrictions can be reduced by changing the $\tau_{\Pi}$ parametrization. 
%Finally, they conclude that a 20 \% theoretical uncertainty can emerge in Bayesian inferences in small systems (ultraperipheral ion-ion collisions and also high multiplicity proton-ion collisions) by regulating if the simulations are regulated not to violate causality.  
%Besides that, the fact that ugly ($w^{2} < 0$) regions are contained within the bad or the ugly ($v w > 1$) regions, discussed in the context of this work's bulk-only test, is not general. 

\section{Conclusion}
\label{sec:concl}

In the present work, we investigated the interplay between causality, instability through anti-dissipation, and numerical solutions of relativistic viscous hydrodynamic equations.
 For the sake of simplicity, we considered systems that contain only the bulk viscous stress. In this case, causality is valid if the fluid's characteristic propagation of information speed $w$ is real and subluminal. In case $w$ is superluminal, these violations of causality only lead to instability due to anti-dissipation if $vw > 1$; then, as argued in Section~\ref{hy2}, numerical instabilities are also expected. Another pathological regime occurs when $w^{2} < 0$; then, in the mathematical classification, the equations of motion become elliptic and a numerically unstable behavior of the simulations is also expected. 
 
 As a controlled test, we considered a space-homogeneous system with only bulk viscous stress and constant speed of sound. In that case, an analytical evolution can be derived for the bulk stress and the velocity of the fluid. This serves as a benchmark for the expected behavior of hydrodynamic simulations, since the causal and stable (`good'), acausal and stable (`bad'), and acausal and unstable (`ugly') regimes have straightforward definitions. In this analytical example, one can set initial conditions in these different regimes, and observe the behavior of the particular hydrodynamic numerical solver.
We found that the hydrodynamic solver MUSIC recovered our ($0+1$)D  benchmark solution in the causal (`good'), and the acausal but stable (`bad') cases, as long as implementation-specific fixes were applied to the initialization of time derivatives in the solver (see Section~\ref{sec:num-sol-bulk}).   
On the other hand, a numerical runaway solution was observed in the unstable (`ugly') case. This solution does not coincide with the analytical solution, as expected from the accumulation of errors typical of this regime (see Section~\ref{hy1}). Hence, only unstable (`ugly') initial conditions lead to significant numerical inconsistencies. 

As discussed in Section~\ref{eq:GBU-test-2D}, the coordinate system employed by viscous hydrodynamic solvers in heavy-ion physics means that the identified time derivative initialization issue --- common among similar Eulerian solvers in the field --- will not necessarily have phenomenological implications. Nevertheless, this finding underscores the value of having access to a broad range of analytical solutions for validating numerical solvers.

For $(2+1)$D hydrodynamic evolution with an equation of state and initial conditions relevant to relativistic heavy-ion collisions, 
we saw that a minimal Israel-Stewart model can produce acausal and unstable regions even for causal initial states. Acausal regions tend to occur near some local wavefronts, where the system departs significantly from equilibrium. 
On the other hand, incorporating a single second-order term ($\delta_{\Pi \Pi} \Pi \partial_\mu u^\mu $) almost completely eliminated acausal and unstable regions, while producing a very similar temperature profile. Evidently, one should not necessarily include second-order terms selectively without a guiding principle. However, our results do suggest that stability and causality analysis may depend significantly on second-order terms, while the hydrodynamic evolution itself may be much less sensitive to their inclusion.

Acausal behavior has been documented in prior work~\cite{Plumberg:2021bme,Chiu:2021muk,ExTrEMe:2023nhy,Domingues:2024pom} for simulations that incorporate shear viscosity or shear-bulk coupling terms. It will be valuable to understand if instabilities are associated with the observed acausalities. Additionally, it will be important to investigate how localized instabilities propagate and influence the global solution behavior, and whether numerical implementations may inadvertently regulate these instabilities through discretization schemes or other computational effects.

We remark that the pathologies discussed here are guaranteed to appear in all versions of Israel-Stewart theory (rBRSSS~\cite{Baier:2007ix},  DNMR~\cite{Denicol:2012cn}, IReD~\cite{Wagner:2022ayd}), if pushed sufficiently far away from equilibrium. Indeed, they are expected to appear in all transient hydrodynamic theories (i.e. in all theories where the non-equilibrium fluxes are promoted to degrees of freedom). This includes, for example, Divergence-Type Theories (DTTs) \cite{GerochDivergence}. In fact, while in principle DTTs are capable of remaining causal far from equilibrium, this can only happen if the generating functional remains well-behaved at arbitrarily large values of the dissipation tensor. Unfortunately, no such well-constructed generating functional has ever been found.
By contrast, BDNK \cite{Bemfica:2017wps, Bemfica:2019knx, Bemfica:2020zjp, Kovtun:2019hdm} (which is not a transient theory) is completely immune to these pathologies, since its causality properties do not depend on how far from equilibrium the system is (i.e. on how large gradients the fluid is experiencing). Hence, a well-constructed BDNK theory will always be causal and well-posed.

\section*{Acknowledgments}

The authors thank Dekrayat Almaalol, Fábio S.~Bemfica, Jacquelyn Noronha-Hostler, Matthew Luzum and Christopher Plumberg for useful discussions. H.H., J.-F.~P., M.~S. and G.~S.~R. are supported by Vanderbilt University and by the U.S. Department of Energy, Office of Science under Award Number DE-SC-0024347.
H.H. is also partly supported by the U.S. Department of Energy, Office of Science under Award Number DE-SC0024711, and the National Science Foundation under Grant No. DMS-2406870. L.G. is partially supported by a Vanderbilt Seeding Success grant.

\appendix

\section{Homogeneous fluid with only bulk stress: obedience to the conservation laws}
\label{sec:hom-fluid-obed}

\begin{figure}[!h]
    \centering
    \includegraphics[scale=0.4]{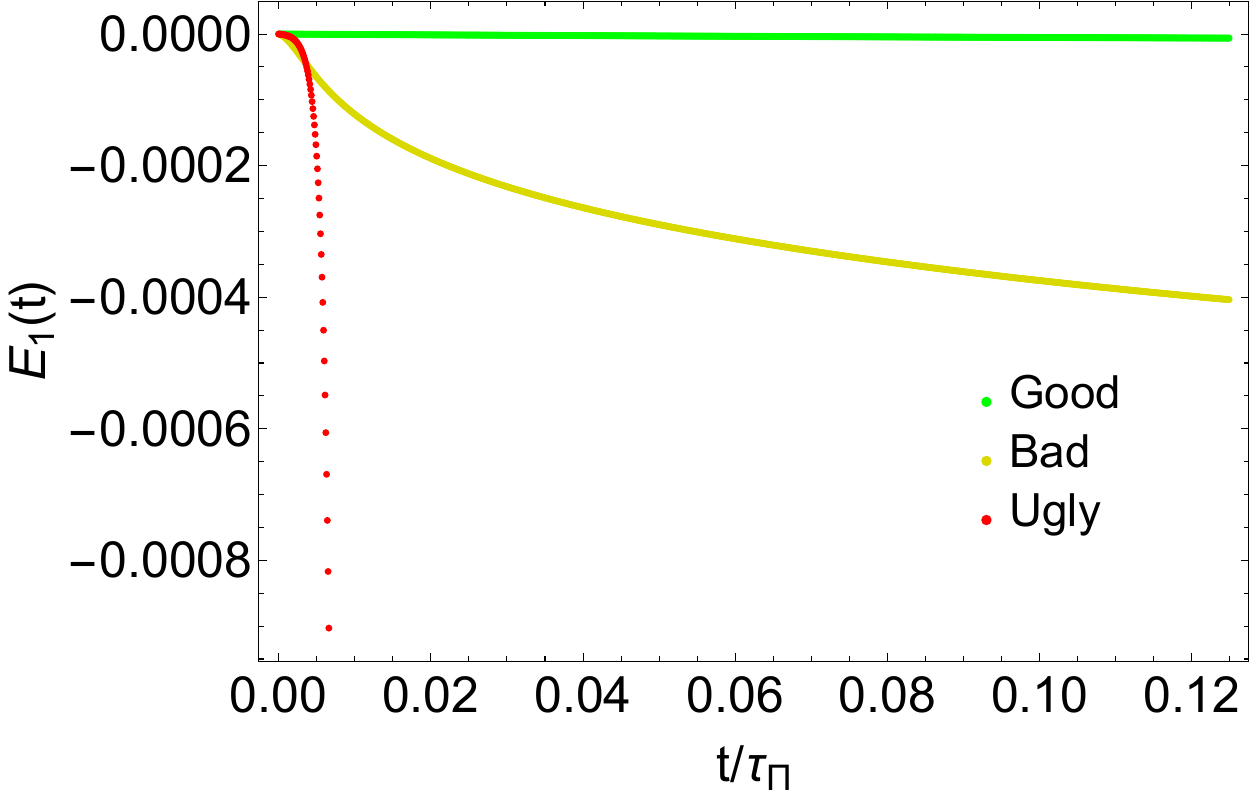}
    \includegraphics[scale=0.4]{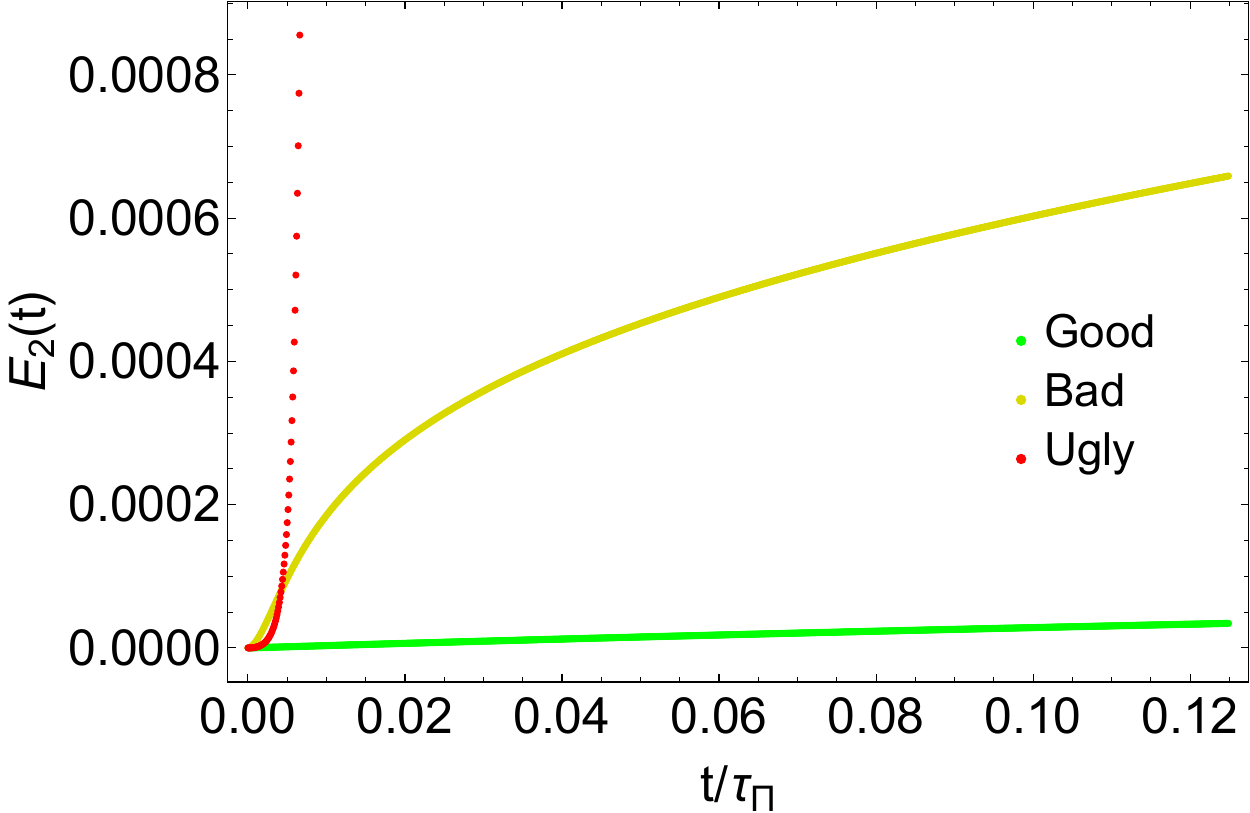}
\caption{Error estimators as a function of normalized time for simulations with analytical pre-initial conditions. (Left panel) $E_{1}(t)$ (see eq.~\eqref{eq:e1}). (Right panel) $E_{2}(t)$ (see eq.~\eqref{eq:e2}).}
\label{fig:E12-GBU-AA}
\end{figure}

\begin{figure}[!h]
    \centering
    \includegraphics[scale=0.4]{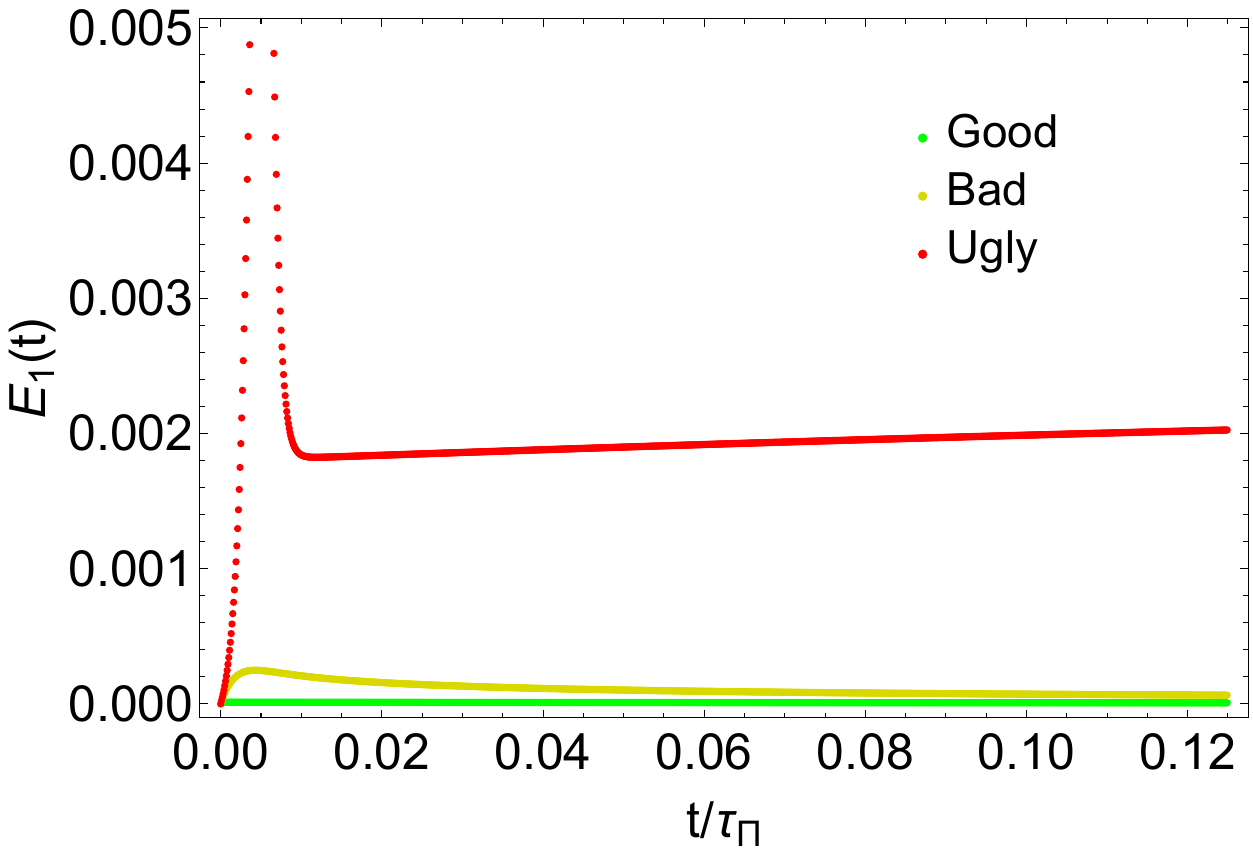}
    \includegraphics[scale=0.4]{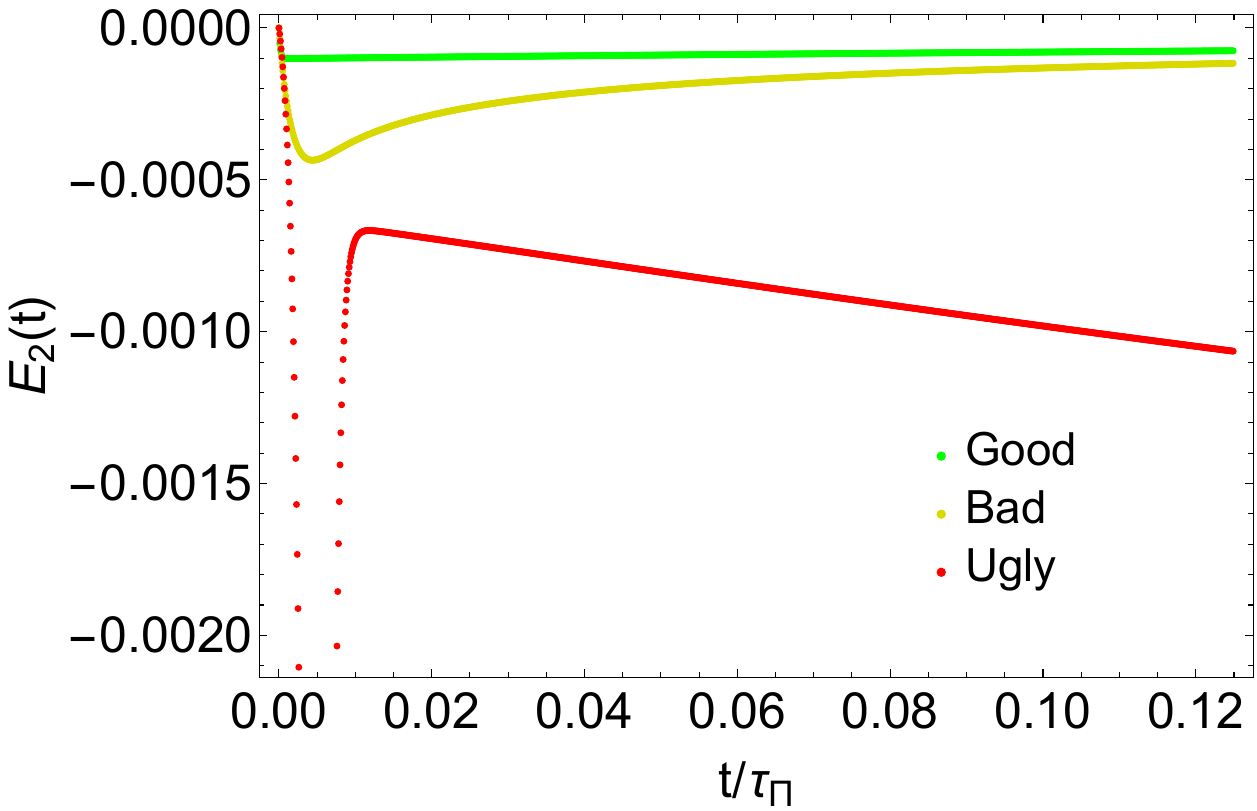}
\caption{Error estimators as a function of normalized time for simulations with zero-time-gradient initialization. (Left panel) $E_{1}(t)$ (see eq.~\eqref{eq:e1}). (Right panel) $E_{2}(t)$ (see eq.~\eqref{eq:e2}).}
\label{fig:E12-GBU-ZG}
\end{figure}

It was mentioned in Section~\ref{sec:num-sol-bulk} that numerical solutions to the hydrodynamic equations of motion, for good and bad initial conditions, obey the local conservation laws in eq.~\eqref{garzone}. The goal of this Appendix is to check this explicitly. To do this, let us note that eq.~\eqref{garzone} relates the thermodynamic pressure and the bulk viscous pressure at a given time $t$ to their corresponding counterparts at the initial time as follows:
\begin{equation}
\begin{aligned}
&
\frac{P(t)}{P(0)} = \frac{\mathcal{E}/\mathcal{P}^1- v(t)}{\mathcal{E}/\mathcal{P}^1- v(0)},
\\
&
\frac{\Pi(t)}{\Pi(0)} 
=
\frac{\dfrac{1}{v(t)}+ c_s^2 v(t) -(1{+}c_s^2)(\mathcal{E}/\mathcal{P}^1)}{\dfrac{1}{v(0)}+ c_s^2 v(0) -(1{+}c_s^2)(\mathcal{E}/\mathcal{P}^1)}.
\end{aligned}    
\end{equation}
In order to assess violations to the expressions above, and thus to the conservation laws, we shall employ the following relative error functions estimators
\begin{subequations}
\begin{align}
&
\label{eq:e1}
E_{1}(t) = \left( 
\frac{P(t)}{P(0)} - \frac{\mathcal{E}/\mathcal{P}^1- v(t)}{\mathcal{E}/\mathcal{P}^1- v(0)}
\right)\frac{P(0)}{P(t)}
\\
&
\label{eq:e2}
E_{2}(t) = \left( \frac{\Pi(t)}{\Pi(0)} 
-
\frac{\dfrac{1}{v(t)}+ c_s^2 v(t) -(1{+}c_s^2)(\mathcal{E}/\mathcal{P}^1)}{\dfrac{1}{v(0)}+ c_s^2 v(0) -(1{+}c_s^2)(\mathcal{E}/\mathcal{P}^1)}
\right) \frac{\Pi(0)}{\Pi(t)},
\end{align}    
\end{subequations}
which are displayed in figs.~\ref{fig:E12-GBU-AA} and \ref{fig:E12-GBU-ZG} below for the three causality-stability statuses. They should be such that $E_{1,2}(t) \approx 0$ at all times. In what follows, we shall employ $\{c_s^2,a,\tau_\Pi,\mathcal{E}/\mathcal{P}^1 \}=\{1/3,1,1 \text{ fm},2 \}$, as in the main text. 

In fig.~\ref{fig:E12-GBU-AA}, we display the time evolution of $E_{1}(t)$ and $E_{2}(t)$ considering the analytical initial state (corresponding to the left panel of  fig.~\ref{fig:GBU-preIC} in the main text). We readily see that both error functions are largest in magnitude for `ugly' initial conditions, followed by `bad' and `good'. For `good' initial conditions, there is a very slow (within the scale analyzed) increase in the behavior of both error functions with time in a way that both functions remain an order of magnitude below the equivalent estimates for the other types of initial conditions in fig.~\ref{fig:E12-GBU-ZG}. For the `bad' initial conditions, we see a stronger increase in the accumulation of errors with time, but it does not surpass 0.06\% due to the fine timestep discretization. As expected, for the `ugly' initial conditions, we see a very steep increase in the error functions which is another signal of the runaway behavior seen in fig.~\ref{fig:GBU-preIC} (right panel). At $t/\tau_{\Pi} \approx 0.015$, the runaway numerical solution approaches $v(t) \approx 1$ and the error estimators are much larger than the in the `good' and `bad' counterparts.\footnote{When the numerical regulator mentioned in Footnote \ref{fn:QRV-MUSIC} is turned on, after $v(t) \approx 1$, the numerical code artificially sets $v(t) \approx 0$. This creates another discontinuity; then the error functions behave so that $E_{1}(t) \sim - 10^{16}$ and $E_{2}(t) \sim 10^{23}$}

In fig.~\ref{fig:E12-GBU-ZG}, we show the numerical error functions considering the `zero-time-gradient' initial condition, as explained in Section~\ref{sec:num-sol-bulk} (corresponding to the right panel of  fig.~\ref{fig:GBU-preIC} in the main text). There, we see that the estimators still follow the same ordering in magnitude as in the analytical initial conditions described above, but with a very different overall evolution. In both $E_{1}$ and $E_{2}$, the quick `jump' from the unstable to the stable branch at $t/\tau_{\Pi} \sim 0.05$ (see fig.~\ref{fig:GBU-preIC}, right panel) for `ugly' initial conditions manifests itself as a very narrow peak. For $t/\tau_{\Pi} \gtrsim 0.05$ the error estimators decrease in absolute value and show a slow linear growth, indicating that the conservation laws are being obeyed. This, however, occurs in a manner that is inconsistent with the initial conditions. Practically an order of magnitude below that, the error functions for `bad' initial conditions have a broadly peaked curve with a maximum also around $t/\tau_{\Pi} \sim 0.05$, after which the functions decrease in magnitude. As for `good' initial conditions, the error functions range much smaller values than the previous two regimes, such that no curve structure is resolvable in the scale where the plots are made.

\newpage

\section{Details on numerical initial condition prescriptions in hydrodynamic simulations}
\label{apn:details-IC}

In Section~\ref{sec:num-sol-bulk}, we discussed the zero-time-gradient initial conditions employed in solvers of relativistic dissipative hydrodynamic equations. In this Appendix, since this approach introduces spurious numerical artifacts, the goal is to outline a general iterative numerical prescription to solve for the initial time step gradients. As in the main text, we shall consider a system with only bulk viscosity, but analogous arguments should be enough for more generic fluid configurations. Once again, our starting point is the local conservation laws and the relaxation equation for the bulk viscous pressure expressed in the form
\begin{align}
    \partial_t T^{t\mu}_{\rm id} = -\partial_i T^{i\mu}_{\rm id} - \partial_\nu T_{\rm diss}^{\nu\mu}\label{eq:idealevolution}\\
    \partial_t\Pi = -\frac{1}{\gamma\tau_\Pi}\left(\Pi + \zeta\partial_{\mu}u^{\mu}\right).\label{eq:bulkevolution}
\end{align}
where $T^{\mu\nu}_{\rm id} = \varepsilon u^{\mu}u^{\mu} + P(\varepsilon)(g^{\mu \nu} +  u^\mu u^\nu)$ and $T_{\rm diss}^{\mu\nu} = \Pi(g^{\mu\nu} + u^\mu u^\nu)$. As discussed in the main text, the right-hand sides of these equations are considered as ``source terms'' that are evaluated first at each time step. However, since these sources contain time derivatives themselves, a prescription is needed to compute these derivatives at the initial time step. A common assumption is to consider these derivatives to be zero at initial time. This can also be stated as the assumption that $(\varepsilon, u^{\mu}, \Pi)\vert_{t = t_0} = (\varepsilon, u^{\mu}, \Pi)\vert_{t = t_0 - \delta t}$, where $ t = t_0-\delta t$ is some fictitious pre-initial time.

Since the source terms approach leads to significant numerical simplicity, a possible solution to keep using it is the following.  
We begin with an initial guess for the field values at the fictitious time $t = t_0 - \delta t$. Then eqs. (\ref{eq:idealevolution}-\ref{eq:bulkevolution}) can be solved. We can then use this solution to adjust the field values at $t = t_0 - \delta t$. We repeat this procedure iteratively until we find values for all temporal derivatives of interest that match the ones computed numerically after the evolving one time step. Once the temporal derivatives are found for the first time step, we move ahead with the conventional temporal evolution.

In our simple example presented in Section~\ref{sec:num-sol-bulk}, the system seems to approach the analytic solution very quickly, if the system is in the causal (`good') or the stable acausal (`bad') regime, even in the absence of a proper solution for the first time step. For the unstable regime, the zero temporal derivative assumption acts as a regulator and effectively kills the unstable modes, and the system lands in some smooth solution, albeit not the correct unstable one. As discussed in Section~\ref{eq:GBU-test-2D}, in Milne coordinates, the effect can become subdominant if sufficiently small hyperbolic time $\tau$ is employed for the initial conditions.

\bibliography{liography}
\bibliographystyle{apsrev4-2}

\end{document}